\documentclass[draft,12pt]{elsarticle}
\usepackage{subfigure}
\usepackage{amssymb}
\usepackage{stmaryrd}
\usepackage{mathrsfs}
\usepackage{amsmath}
\usepackage{amssymb}
\usepackage{multirow}

\newcommand{\test}{\mbox{$
\begin{array}{c}
\stackrel{ \stackrel{\textstyle H_1}{\textstyle >} } {
\stackrel{\textstyle <}{\textstyle H_0} }
\end{array}
$}}
\begin{document}

\begin{frontmatter}

\title{WRFRFT-based Coherent Detection and Parameter Estimation of Radar Moving
Target With Unknown Entry/Departure Time}

\author{Xiaolong L$\text{i}^{*}$, Zhi Sun, Tianxian Zhang, Wei Yi, Guolong Cui, and Lingjiang Kong}

\address{School of Information and Communication Engineering\\
University of Electronic Science and Technology of China\\
Chengdu, Sichuan, P.R. China, 611731\\
Corresponding author: Xiaolong Li\\
Email: xiaolongliuestc@gmail.com}

\begin{abstract}
A moving target may enter a radar coverage area unannounced and leave after an unspecified period, which implies that the target's entry time and departure time are unknown. In the absence of these time information, target detection and parameter estimation (DAPE) will be severely impacted.
In this paper, we consider the coherent detection and parameters estimation
problem for a radar moving target with unknown
entry time and departure time (that is,
the time when the target appears-in/leaves the radar detection field
is unknown), involving across range cell (ARC) and Doppler spread (DS)
effects within the observation period.
A new algorithm, known as window Radon Fractional
Fourier transform (WRFRFT) is proposed to
detect and estimate the target's time
parameters (i.e., entry time and departure time)
and motion parameters (i.e., range, velocity and
acceleration). The observation values of a maneuvering target are first
intercepted and extracted by the window function and
searching along the motion trajectory. Then these
values are fractional
Fourier transformed and well accumulated in the WRFRFT domain,
where the DAPE of target could be
accomplished thereafter.
 Experiments with simulated
and real radar data sets prove its effectiveness.

\end{abstract}

\end{frontmatter}



\section{Introduction}\label{sect: introduction}

Radar is widely used in military and civilian surveillance  because of its  ability of working at all-time and under all-weather conditions \cite{DPTsp2018,TRTsp2017,SunC2015SP,ZhongH2010, ZhengJB001,HPHstkcpf2017}.
Generally speaking, target detection, parameter estimation and target imaging are the three most important functions of modern radars \cite{ZhengJB003,ZhengJB016ST,Liieeesj02}, where the performance of target detection and parameter estimation will significantly affect the imaging quality for the target \cite{RFRAF,Zhu00,Tian}.  Therefore, how to improve the performances of  target detection and parameter estimation have become important topics in  radar research \cite{ChenXL01,HuangPnew02,XuJiaaesM02}.

Long-time coherent accumulation is an effective method to improve the performance of radar target detection and parameter estimation (DAPE) \cite{Huang1}. By stacking the echo signals in the same phase
 level within a certain observation time, the echo's signal-to-noise ratio (SNR) can be improved to the maximum extent with coherent integration processing \cite{aes2}. However, due to the movement of the target, the phenomenon of across-range cell
(ARC) and Doppler spread (DS) \cite{XuJiaaesM02} will occur within the accumulation time, leading to the failure of the traditional moving target detection (MTD) method. More importantly, in the actual detection scene, the target's time parameters information (that is, when the target enters and leaves the radar detection area) and motion parameters information (i.e.,
distance, velocity, and acceleration) are often unknown, which will bring great challenges to the DAPE of moving targets with long time coherent integration.

To deal with the ARC and realize the coherent detection of constant velocity
targets, many methods have been proposed, such as location rotation transform
(LRT) \cite{Sunjrs2018}, sequence reversing transform (SRT) \cite{LiXL2017SRT},
Radon Fourier transform (RFT) \cite{Xu14,Xub15,Yu16},
scaled inverse Fourier transform (SIFT) \cite{ZhengJB002} and keystone
transform (KT) \cite{Perry06,Li,ZhuDY}. More specifically, LRT and RFT eliminate the ARC by two-dimensional
searching processing in the parameter space, while SRT and SIFT are able to
eliminate the ARC via the correlation operation
without the brute-force searching process.
Unfortunately, when the moving target has a certain degree of mobility
(e.g., with acceleration), the methods in \cite{Sunjrs2018,LiXL2017SRT,Xu14,Xub15,Yu16,ZhengJB002,Perry06,Li,ZhuDY}
may suffer integration and
detection performance loss since they could not remove the
ARC and DS caused by target's acceleration motion.

As to the accelerating target, it is worth noting that the ARC (including the
ARC induced by velocity and the ARC caused by
acceleration) and DS effects occur simultaneously during the observation time.
To jointly correct and compensate for the ARC and DS,
Xu, et al. \cite{Xu17} introduced a method known as generalized
Radon Fourier transform (GRFT),
which is the extension of the original RFT algorithm.
In addition, Li, et al. \cite{LiXLRLVD} proposed an algorithm named Radon Lv's distribution (RLVD) to accumulate and detect
weak target signals, while Chen, et al. \cite{ChenXL} proposed the Radon Fractional Fourier
transform (RFRFT) for a maneuvering target detection. To some extent, these two algorithms
\cite{LiXLRLVD,ChenXL} also can be regarded as the expansion and extension of the RFT, since
both RLVD and RFRFT  eliminate the ARC and DS simultaneously by searching
in the range-velocity-acceleration domain.
However,
the computational complexity of RLVD/RFRFT is bigger than that of
GRFT.

Moreover, there are some other coherent integration algorithms, which could
be employed to deal with the ARC and/or DS, such as short time GRFT (STGRFT) \cite{LiXLtspSTGRFT},
symmetric autocorrelation function \cite{SPsafsft},
adjacent correction function \cite{LiXL2015,LiXLtsp2016,LXLsj2015accf},
KT with matched filter process \cite{SPktmfp}, three-dimensional scaled transform
\cite{ZhengJB18}, and so on.
In particular, the STGRFT-based method is able to remove the ARC/DS and obtain the
coherent integration.
However, the STGRFT-based method only considers the detection problem
of a moving target. Most importantly, STGRFT and the other existing algorithms mentioned above
\cite{Sunjrs2018,LiXL2017SRT,Xu14,Xub15,Yu16,ZhengJB002,Perry06,Li,ZhuDY,Xu17,LiXLRLVD,ChenXL,
LiXLtspSTGRFT,SPsafsft,LiXL2015,LiXLtsp2016,LXLsj2015accf,SPktmfp,ZhengJB18}
are all
assume that the target's time parameter information (that is,
the time when the target entries/leaves the radar detection
area) is already known.
Actually, before the
target detection and parameters estimation are accomplished,
the time of target's entry and leaving is  often unknown,
and then this assumption will no longer hold and
the existed coherent integration methods would become
invalid, or at least degraded.

This paper addresses the coherent detection and parameters estimation
problem for a radar moving target with unknown entry/departure time,
where the ARC and DS effects are also considered.
A new method known as window Radon Fractional Fourier transform (WRFRFT) is
proposed to estimate the target's time parameters (i.e., entry time and
departure time)
and motion parameters (i.e., range, velocity and acceleration).
By employing the window function and searching process within the parameter space,
the WRFRFT is capable to remove the ARC/DS effect and realize the
coherently refoucusing of target signal, resulting in superior SNR improvement and
better detection and parameter estimation ability.
Detailed experiments with simulated data and real radar data are given to
demonstrate the superiority of the WRFRFT-based method.

The signal model for a radar moving target with unknown entry time and
departure time is
described in Section \ref{sect: signal model}.
 In section \ref{sec:WRFRFT}, the proposed WRFRFT method is given, including
its definition, properties, main steps and computational complexity.
In Section \ref{sec:experiments}, the performances of WRFRFT method
with simulated and real data
are assessed.
Section \ref{sec:conclusion} presents the conclusions.

\section{Mathematical Model of Received Signal }\label{sect: signal model}

Assume that the linear frequency modulation is adopted as
the radar's transmitted waveform, i.e.,
%
\begin{equation}\label{eq1}
\begin{split}
s_{trans}(\hat{t})=\text{rect}\left(\frac{\hat{t}}{T_p}\right)\exp\left(j\pi \gamma\hat{t}^2\right)\exp(j2\pi
f_c \hat{t})
\end{split}
\end{equation}
where $
 \text{rect}\left( x
\right)=\left\{
\begin{matrix}
   1, & \left| x \right|\le 0.5  \\\nonumber
   0, & \left| x \right|> 0.5  \\
\end{matrix} \right.
$,
$\hat{t}$, $\gamma$, ${{f}_{c}}$ and
$T_p$ denote, respectively, the fast time variable,
chirp rate, carrier frequency and pulse duration.

Suppose that the total observation time of radar is from $T_0$ to $T_1$,
where a moving target enters the radar detection area
at time $T_b$ and leaves the radar detection area
at time $T_e$ ($T_0 < T_b <T_e < T_1$).
The instantaneous slant distance between the target and radar at
time $T_b$ is denoted as $R_0$,
and the radial distance of target could be expressed as:
\begin{equation}
R(t)=R_{0}+V(t-T_b)+A(t-T_b)^2,    t\in [T_b, T_e]
\end{equation}
where $t$ denotes the slow time, while $V$ and $A$ are respectively the
target's radial velocity and acceleration,
$T_b$ represents the  beginning-time (entry) of
the target  and  $T_e$  represents the ending-time (departure) of
the target, which are both unknown.

With the pulse compression (PC),
the received signal within the observation time
can be expressed as \cite{Xu17}
\begin{equation}\label{eq3}
\begin{split}
s(\hat{t},t)&=w(t) \sigma_0\text{sinc}\left[B\left(\hat{t}-\frac{2R(t)}{c}\right)\right]\\
&\times\exp\left[-j4\pi\frac{R(t)}{\lambda}\right]+n_s(\hat{t},t)
\end{split}
\end{equation}
where
\begin{equation}\label{eq4}
\begin{split}
w(t)=\text{rect}\left[\frac{t-0.5(T_b+T_e)}{T_e-T_b} \right]
=\left\{
\begin{matrix}
   1, & T_b\leq t\leq T_e  \\
   0, & \text{else}  \\
\end{matrix} \right.
\end{split}
\end{equation}
$n_s(\hat{t},t)$
represents noise, $\sigma_0$ and $\lambda$ are
respectively the signal amplitude and the wavelength, while
$B$ and $c$ represent the bandwidth and light speed, respectively.

\begin{figure}[!htbp]
\begin{center}
\includegraphics[width=1 \textwidth,draft=false]{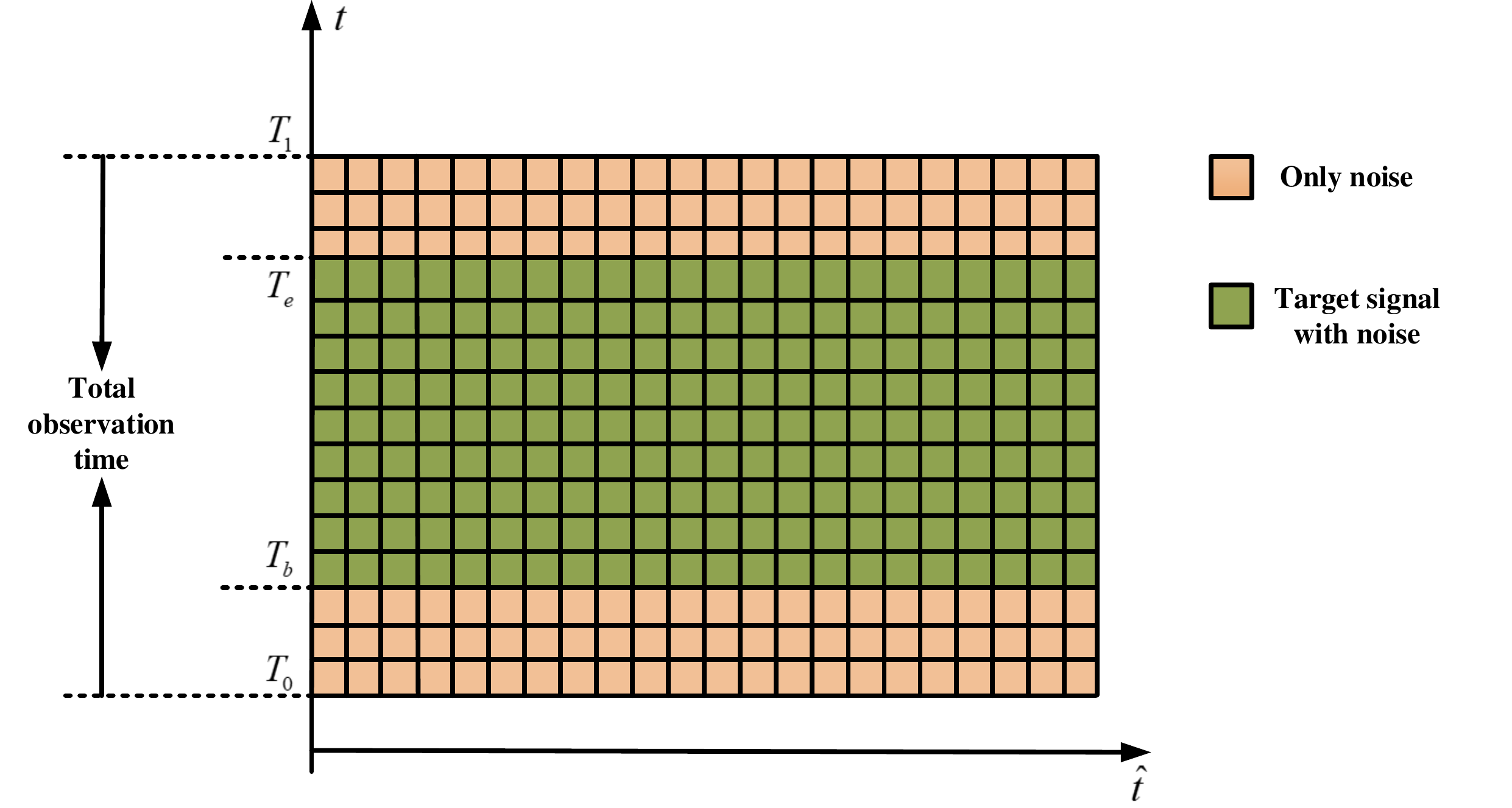}
\end{center}
\caption{Sketch map of the radar's echo  in the $\hat{t}$-$t$ plane.}\label{fig:sketchmap}
\end{figure}

From \eqref{eq3}, it could be noticed that the radar echo contains target signal
within the period $[T_b, T_e]$, while the radar echo only contains noise for the other
time periods, i.e., the beginning time of target signal is $T_b$ and the ending time
of target signal is $T_e$.
The sketch map of the radar echo in the $\hat{t}$-$t$ plane is  given
in Fig. \ref{fig:sketchmap}.

\section{WRFRFT-based Method}\label{sec:WRFRFT}

\subsection{Definition of WRFRFT}

The definition of WRFRFT is given as
\begin{equation}\label{eq5}
\begin{split}
WR_{g(t)}(\alpha,u)&=F_{\alpha}[g(t)s\left(2r(t)/c,t\right)](u)\\
&=\int_{-\infty}^{\infty}g(t)s\left(2r(t)/c,t\right)K_a(t,u)dt,
\end{split}
\end{equation}
where $g(t)$ and $r(t)$ are respectively the window function and  searching
motion trajectory:
\begin{equation}\label{eq6}
r(t)=r_{0}+v(t-\eta_0)+a(t-\eta_0)^2
\end{equation}
\begin{equation}\label{eq7}
\begin{split}
g(t)=&\text{rect}\left[\frac{t-0.5(\eta_{1}+\eta_{0})}{\eta_{1}-\eta_{0}} \right]
=\left\{
\begin{matrix}
   1, & \eta_{0}\leq t\leq \eta_{1}  \\
   0, & \text{else}  \\
\end{matrix} \right.
\\
&
T_0 \leq \eta_0 \leq \eta_1 \leq T_1
\end{split}
\end{equation}
$\eta_0$ is the beginning time of
$g(t)$
and $\eta_1$ is ending time of $g(t)$, $(r_0,v,a)$ denotes the searching motion
parameters pair (i.e., searching initial range, searching
radial velocity and radial acceleration).
$\alpha=P\pi/2$ is the rotation angle, $P$ is searching transform order,
$F_{\alpha}$ represents the FRFT operator, and the transform
kernel $K_a(t,u)$ is given by
\begin{equation}\label{eq48}
\begin{split}
&K_a(t,u)=\\
&\left\{
\begin{matrix}
   A_{\alpha}\exp[j(0.5t^2\cot\alpha-ut\csc\alpha+0.5u^2\cot\alpha)] &\alpha\neq n\pi  \\
   \delta[u-(-1)^n t] & \alpha=n\pi  \\
\end{matrix} \right.
\end{split}
\end{equation}
where $A_{\alpha}=\sqrt{(1-j\cot\alpha)/2\pi}$.

Inserting \eqref{eq7}  into \eqref{eq5} yields,
\begin{equation}\label{eqnew99}
\begin{split}
WR_{g(t)}(\alpha,u)=\int_{\eta_0}^{\eta_1}g(t)s\left(2r(t)/c,t\right)K_a(t,u)dt
\end{split}
\end{equation}

From the definition  of  WRFRFT,
it could be interpreted as the transform of the
target's intercepted and extracted signal in the FRFT domain.
In particular, the WRFRFT includes three main steps:
The first is an interception of the two-dimensional
compressed signal based on function $g(t)$ (i.e.,
$g(t)$ determines the interception-operation's
beginning/ending time). The second is the
signal extraction process applied to the
intercepted signal, based on the resultant searching motion parameters pair $(r_0,v,a)$.
The third is the FRFT-based integration process.

For comparison, the definition of RFRFT is also given:
\begin{equation}\label{eq9}
\begin{split}
RFRFT(\alpha,u)&=F_{\alpha}[s\left(2r(t)/c,t\right)](u)\\
&=\int_{-\infty}^{\infty}s\left(2r(t)/c,t\right)K_a(t,u)dt\\
&=\int_{T_0}^{T_1}s\left(2r(t)/c,t\right)K_a(t,u)dt
\end{split}
\end{equation}

From \eqref{eqnew99} and \eqref{eq9}, we can notice that both WRFRFT and RFRFT extract
the signal along target's motion trajectory and integrate
with FRFT. The main difference is that RFRFT performs
the extraction and integration operations within the total
observation time, while
the beginning/ending time of WRFRFT is adjustable
(determined by $\eta_0$ and $\eta_1$), thus able to better
 match with the moving target with unknown
entry/departure time.
In particular, the RFRFT can be considered as a special case of
WRFRFT (i.e., when $\eta_0=T_0, \eta_1=T_1$).

\subsection{Some Properties of WRFRFT}

1) Rotational Additivity:
Note that the  kernel of WRFRFT has the
rotational additivity property, i.e.,
\begin{equation}
\int_{-\infty}^{\infty}K_\alpha(t,u)K_\beta(u,z)du=K_{\alpha+\beta}(t,z)
\end{equation}

Thus, it could be easily for us to obtain the
rotational additivity of WRFRFT, i.e.,
\begin{equation}
\begin{split}
&WR_{g(t)}(\alpha+\beta,z)\\
&=F_{\beta}[WR_{g(t)}(\alpha,u)](z)\\
&=\int_{-\infty}^{\infty}K_{\beta}(u,z)
\int_{-\infty}^{\infty}g(t)s\left(2r(t)/c,t\right)K_{\alpha}(t,u)dtdu\\
&=\int_{-\infty}^{\infty}g(t)s\left(2r(t)/c,t\right)
\int_{-\infty}^{\infty}K_{\alpha}(t,u)K_{\beta}(u,z)dudt\\
&=\int_{-\infty}^{\infty}g(t)s\left(2r(t)/c,t\right)K_{\alpha+\beta}(t,z)dt\\
&=F_{\alpha+\beta}[g(t)s\left(2r(t)/c,t\right)](z)
\end{split}
\end{equation}

The rotational additivity of WRFRFT
provides us the solution of the transform between  WRFRFTs
with different transform angles. That is to say,
we only need to
calculate WRFRFT with the total transform order for one
time, which is a major advantage in term of computational efficiency.

2) Inverse WRFRFT (IWRFRFT):
According to the rotational additivity property above,
it also could be concluded that the WRFRFT of angle
$-\alpha$ is the inverse of the WRFRFT with
angle $\alpha$, since that
$F_{-\alpha}(F_{\alpha})=F_{\alpha-\alpha}=F_{0}=I$.
The IWFRFT is
\begin{equation}
g(t)s\left(2r(t)/c,t\right)=\int_{-\infty}^{\infty}WR_{g(t)}(\alpha,u)
K_{-\alpha}(t,u)du
\end{equation}

3) Linear Additivity:
It's easy to find that WRFRFT is linear. In particular,
let $\varepsilon_1$ and $\varepsilon_2$ denote
two constant coefficients, and we have
\begin{equation}
F_{\alpha}[\varepsilon_1x_1+\varepsilon_1x_2](u)=\varepsilon_1F_{\alpha}[x_1](u)+\varepsilon_1F_{\alpha}[x_2](u)
\end{equation}
This property
shows that the WRFRFT satisfies the superposition principle,
which is helpful in the analysis of multi-component signals.

4) Index Commutativity:
Apply \eqref{eq5} for two orders, we have
\begin{equation}
\begin{split}
&F_{\beta}[F_{\alpha}[g(t)s\left(2r(t)/c,t\right)](z)]\\
&=\int_{-\infty}^{\infty}K_{\beta}(u,z)
\int_{-\infty}^{\infty}g(t)s\left(2r(t)/c,t\right)K_{\alpha}(t,u)dtdu\\
&=\int_{-\infty}^{\infty}K_{\alpha}(t,u)
\left[\int_{-\infty}^{\infty}g(t)s\left(2r(t)/c,t\right)K_{\beta}(u,z)du\right]dt\\
&=F_{\alpha}[F_{\beta}[g(t)s\left(2r(t)/c,t\right)](z)]
\end{split}
\end{equation}
Hence, the WRFRFT adheres to the index commutativity property.

5) Parseval Relation:
The WRFRFT also
holds the classical Parseval relation:
\begin{equation}\label{eq16}
\begin{split}
&\int_{-\infty}^{\infty}g(t)x\left(2r_x(t)/c,t\right)y\left(2r_y(t)/c,t\right)dt\\
&=
\int_{-\infty}^{\infty}WR_x\left(\alpha,u\right)WR_y^{*}\left(\alpha,u\right)du
\end{split}
\end{equation}
where $WR_x\left(\alpha,u\right)=F_{\alpha}[g(t)x(t,2r_x(t)/c](u)$,
$WR_y\left(\alpha,u\right)=F_{\alpha}[g(t)y(t,2r_y(t)/c](u)$.
In particular, \eqref{eq16} will  turn into
the energy conservation property when $x=y$, i.e.,
\begin{equation}
\begin{split}
\int_{-\infty}^{\infty}g(t)|x\left(2r_x(t)/c,t\right)|^2dt
=\int_{-\infty}^{\infty}|WR_x\left(\alpha,u\right)|^2du
\end{split}
\end{equation}
The squared magnitude of the WFRFT ($|WR_x\left(\alpha,u\right)|^2$)
thus represent the signal
 energy spectrum  with angle $\alpha$ and window function $g(t)$.

\subsection{WRFRFT for Moving Target Detection and Estimation}

Substitute \eqref{eq3} and \eqref{eq4} into \eqref{eq5} yields,
\begin{equation}\label{eq19}
\begin{split}
WR_{g(t)}(\alpha,u)&=F_{\alpha}[g(t)s\left(2r(t)/c,t\right)](u)\\
&=\int_{-\infty}^{\infty}g(t)w(t)\sigma_0\text{sinc}\left[B\left(
\frac{2r(t)}{c}-\frac{2R(t)}{c}\right)\right]\\
&\times\exp\left[-j4\pi\frac{R(t)}{\lambda}\right]K_{\alpha}(t,u)dt\\
&+\int_{-\infty}^{\infty}g(t)n_s(2r(t)/c,t)K_{\alpha}(t,u)dt
\end{split}
\end{equation}
where the first and the second integral term of \eqref{eq19}
represent respectively the
WRFRFT of  target signal and the WRFRFT of noise.

Let $C$ and $D$ are respectively:
\begin{equation}
\begin{split}
C=\{t|g(t)=1\},
D=\{t|w(t)=1\}
\end{split}
\end{equation}
In other words, $C$ denotes the function
$g(t)$'s non-zero
area while $D$ denotes the function $w(t)$'s
non-zero area.

{\bf Case 1:} When $C\cap D=\varnothing$, we have
\begin{equation}
g(t)w(t)=0
\end{equation}
 In this case,
\eqref{eq19} could be expressed as
\begin{equation}
\begin{split}
WR_{g(t)}(\alpha,u)&=0+\int_{-\infty}^{\infty}g(t)n_s(2r(t)/c,t)K_{\alpha}(t,u)dt\\
&=\int_{\eta_0}^{\eta_1}n_s(2r(t)/c,t)K_{\alpha}(t,u)dt
\end{split}
\end{equation}
Hence, in this case, only  noise is extracted and accumulated,
 but none of the target's signal is extracted and accumulated.

{\bf Case 2:} When $C\cap D\neq\varnothing$, we have
\begin{equation}
\begin{split}
g(t)w(t)
=\left\{
\begin{matrix}
   1, & T'\leq t\leq T''  \\
   0, & \text{else}  \\
\end{matrix} \right.
\end{split}
\end{equation}
where
\begin{equation}\label{eq24}
T'=\text{max}[T_b, \eta_0], T''=\text{min}[T_e, \eta_1]
\end{equation}

In this case, the WRFRFT of \eqref{eq19} can be recast as
\begin{equation}\label{eq25}
\begin{split}
WR_{g(t)}(\alpha,u)
&=\int_{T'}^{T''}\sigma_0\text{sinc}\left[B\left(
\frac{2r(t)}{c}-\frac{2R(t)}{c}\right)\right]\\
&\times\exp\left[-j4\pi\frac{R(t)}{\lambda}\right]K_{\alpha}(t,u)dt\\
&+\int_{\eta_0}^{\eta_1}n_s(2r(t)/c,t)K_{\alpha}(t,u)dt
\end{split}
\end{equation}

From \eqref{eq24} and \eqref{eq25}, we notice that the values of
$\eta_0$ and $\eta_1$ determine the extracted range
of the target signal and noise.
In order to ensure SNR improvement, on the one hand, we need to extract
and integrate all the target signals, and at the same time, we need to extract as little noise as possible.

In particular,
to guarantee that all the target signals are extracted and accumulated,
the following equation should be satisfied
\begin{equation}
T'\leq T_b, T''\geq T_e
\end{equation}

Combining with \eqref{eq24} we have
\begin{equation}\label{eq27}
\eta_0\leq T_b,\eta_1\geq T_e
\end{equation}

In addition to  satisfying the  inequality
 \eqref{eq27}, we also need to ensure that as little noise as possible is extracted.
Thus, we have
\begin{equation}
\eta_0=T_b,\eta_1= T_e
\end{equation}
Then \eqref{eq25} could be rewritten as:
\begin{equation}\label{eq29}
\begin{split}
WR_{g(t)}(\alpha,u)
&=\int_{T_b}^{T_e}\sigma_0\text{sinc}\left[B\left(
\frac{2r(t)}{c}-\frac{2R(t)}{c}\right)\right]\\
&\times\exp\left[-j4\pi\frac{R(t)}{\lambda}\right]K_{\alpha}(t,u)dt\\
&+\int_{T_b}^{T_e}n_s(2r(t)/c,t)K_{\alpha}(t,u)dt\\
&=\int_{T_b}^{T_e}\sigma_0\exp\left[-j4\pi\frac{R(t)}{\lambda}\right]K_{\alpha}(t,u)dt\\
&+\int_{T_b}^{T_e}n_s(2r(t)/c,t)K_{\alpha}(t,u)dt\\
&\ \ \ \ \ \text{when} \ r_0=R_0, v=V, a=A
\end{split}
\end{equation}
Equation \eqref{eq29} shows that the entire target signal is
extracted and could be coherently integrated in the FRFT domain
when the motion parameters of the search match with the motion parameters of target.

Based on the analysis above, it could be noticed that
only when $\eta_0=T_b,\eta_1= T_e$, the
target signal is totally extracted and accumulated
in the FRFT domain while
it could ensure that as little noise as possible is extracted
simultaneously, resulting in a peak value in the
WRFRFT output (where the peak location corresponds to
the time parameters and motion parameters of target).

With different searching
parameters (i.e., beginning/ending time, range, velocity and
acceleration), different integration outputs of
WRFRFT would be obtained and the target signal will be focused as
a peak when the searching time/motion parameters  match with the target's
time/motion parameters.
Thus, the target's time parameters and motion parameters could be estimated by
\begin{equation}
({{\hat{T_b}}},{{\hat{T_e}}},{\hat{R_0}},{\hat{V}},{\hat{A}})
=\underset{{({\eta_0},{\eta_1},{r_0},{v},{a})}}{\arg {\max
}}\,\,|WR_{g(t)}(\alpha,u)|
\end{equation}

\begin{figure*}[!htbp]
\centering \subfigure[]{
\begin{minipage}[b]{0.4\textwidth}\label{fig:different window:a}
\includegraphics[width=1.\textwidth,draft=false]{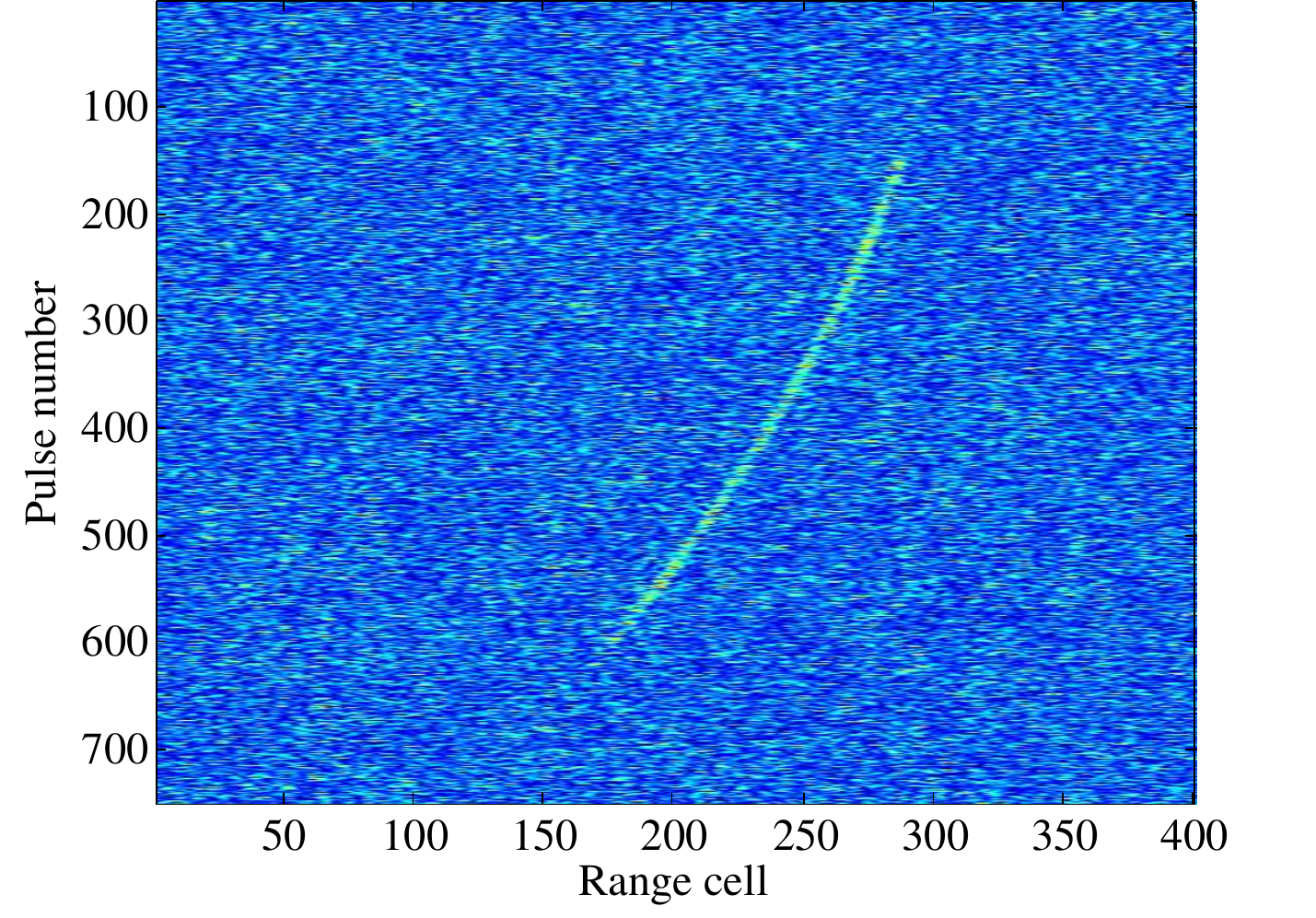}
\end{minipage}
} \subfigure[]{
\begin{minipage}[b]{0.4\textwidth}\label{fig:different window:b}
\includegraphics[width=1.\textwidth,draft=false]{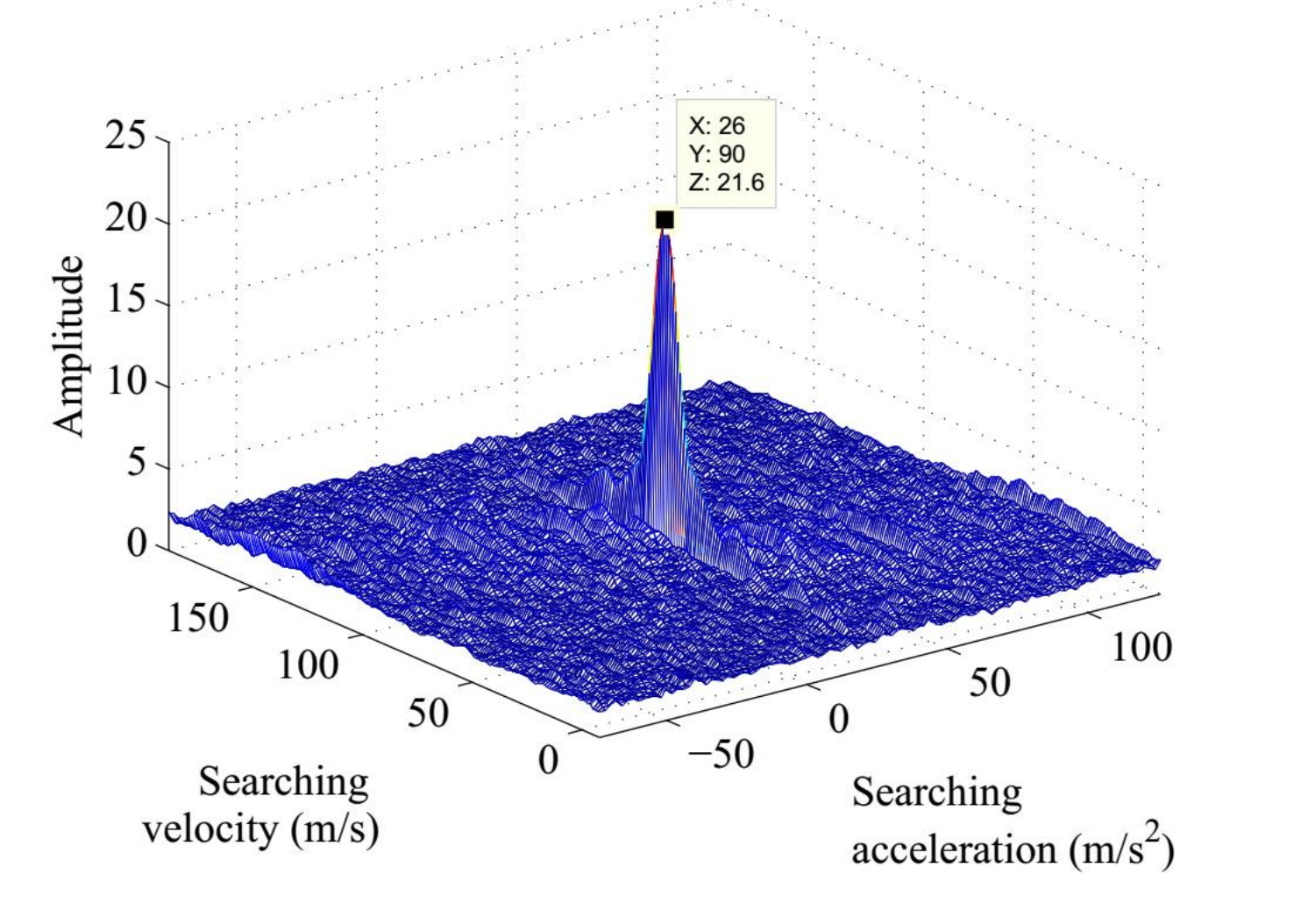}
\end{minipage}
} \subfigure[]{
\begin{minipage}[b]{0.4\textwidth}\label{fig:different window:c}
\includegraphics[width=1.\textwidth,draft=false]{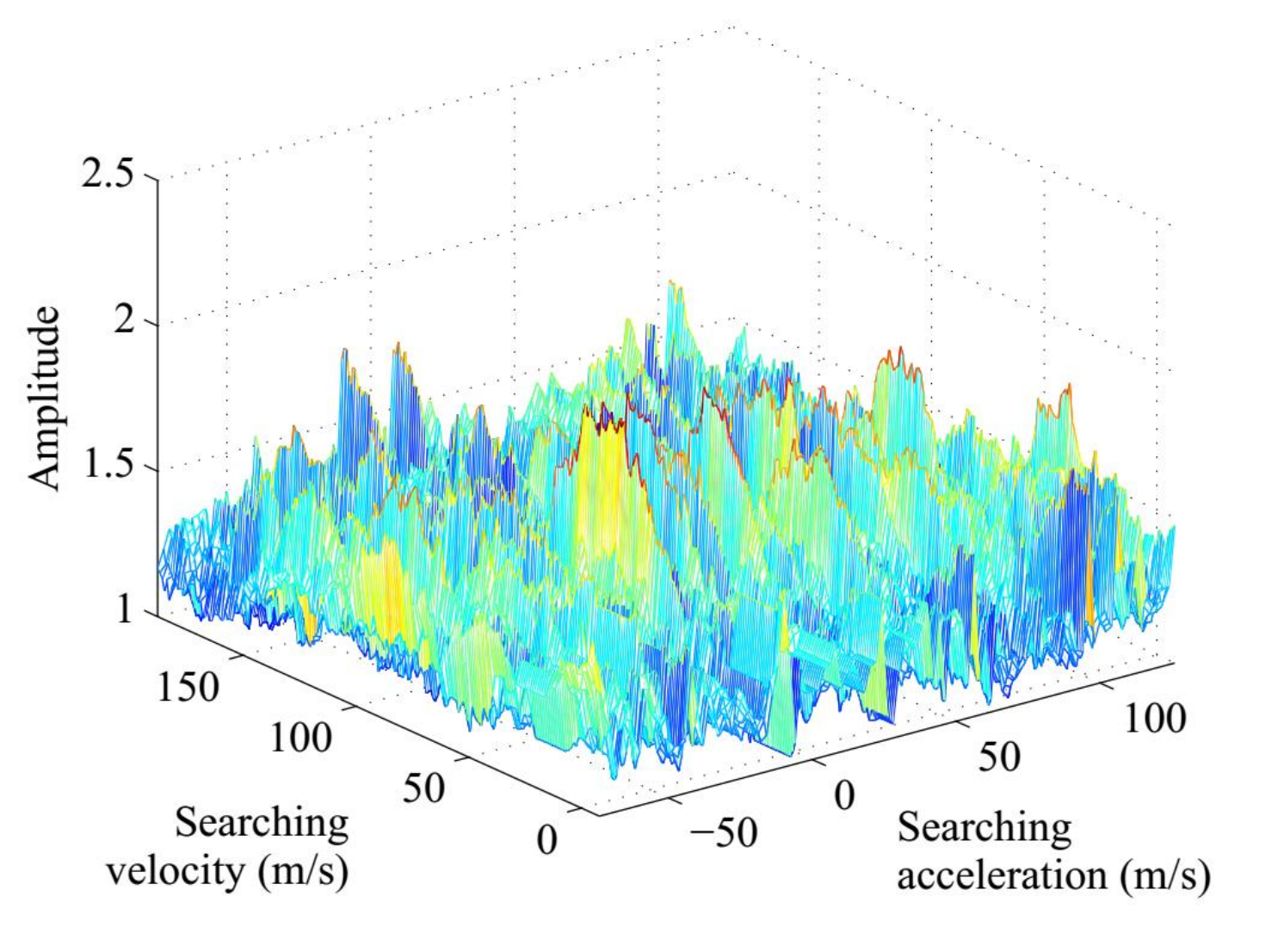}
\end{minipage}
}
\subfigure[]{
\begin{minipage}[b]{0.4\textwidth}\label{fig:different window:d}
\includegraphics[width=1.\textwidth,draft=false]{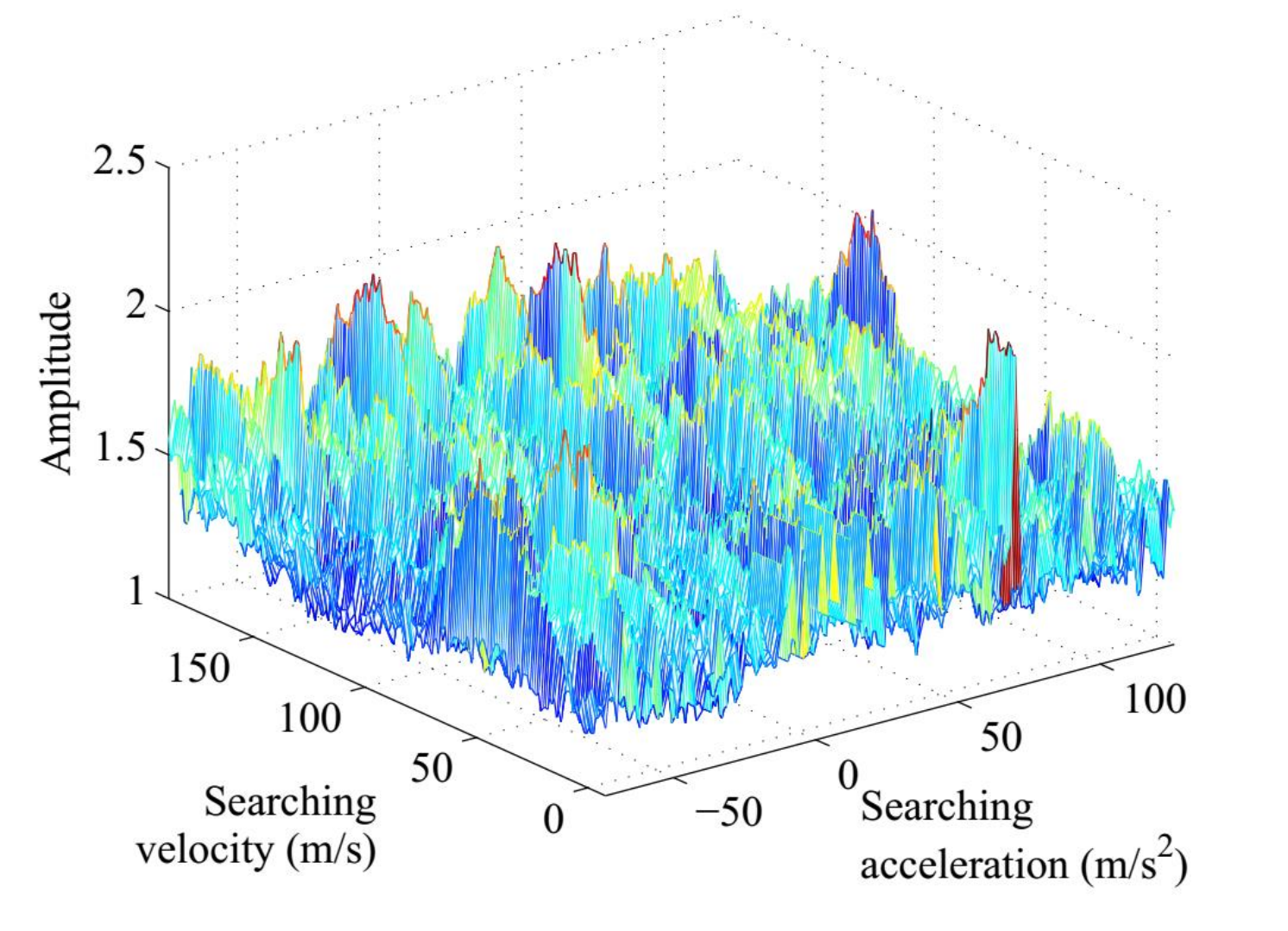}
\end{minipage}
}
\subfigure[]{
\begin{minipage}[b]{0.4\textwidth}\label{fig:different window:e}
\includegraphics[width=1.\textwidth,draft=false]{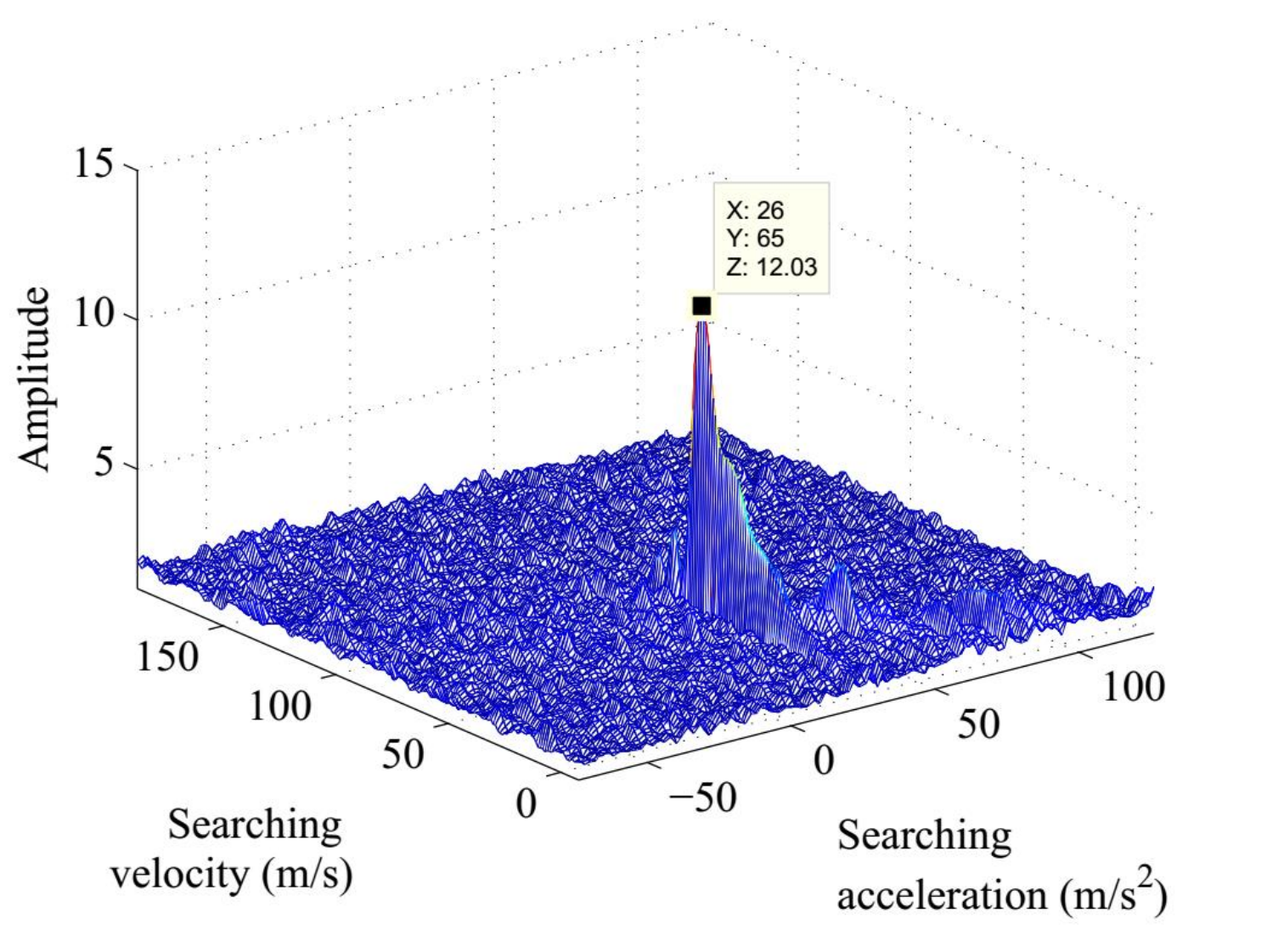}
\end{minipage}
}
\subfigure[]{
\begin{minipage}[b]{0.4\textwidth}\label{fig:different window:f}
\includegraphics[width=1.\textwidth,draft=false]{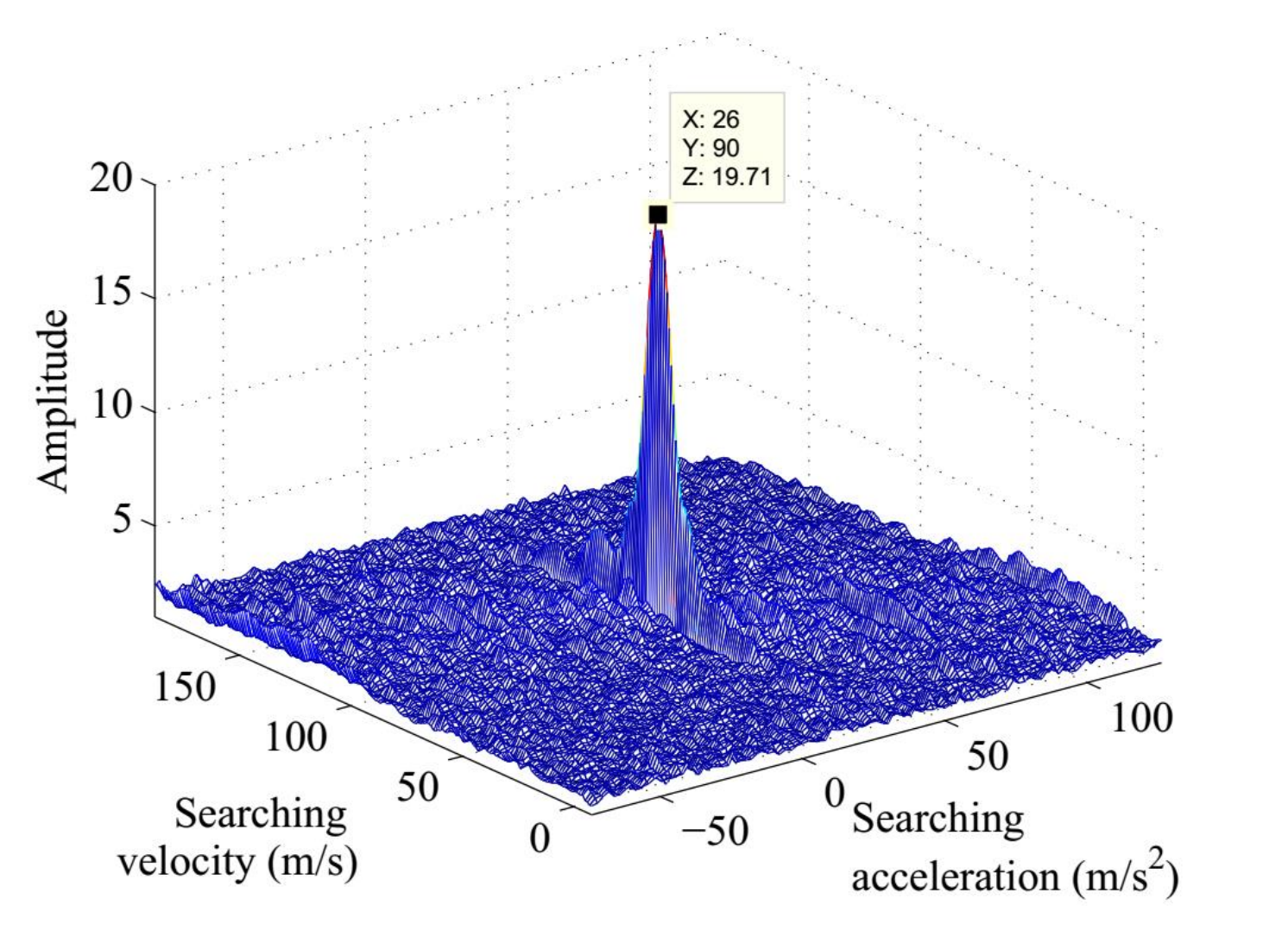}
\end{minipage}
}
\subfigure[]{
\begin{minipage}[b]{0.4\textwidth}\label{fig:different window:g}
\includegraphics[width=1.\textwidth,draft=false]{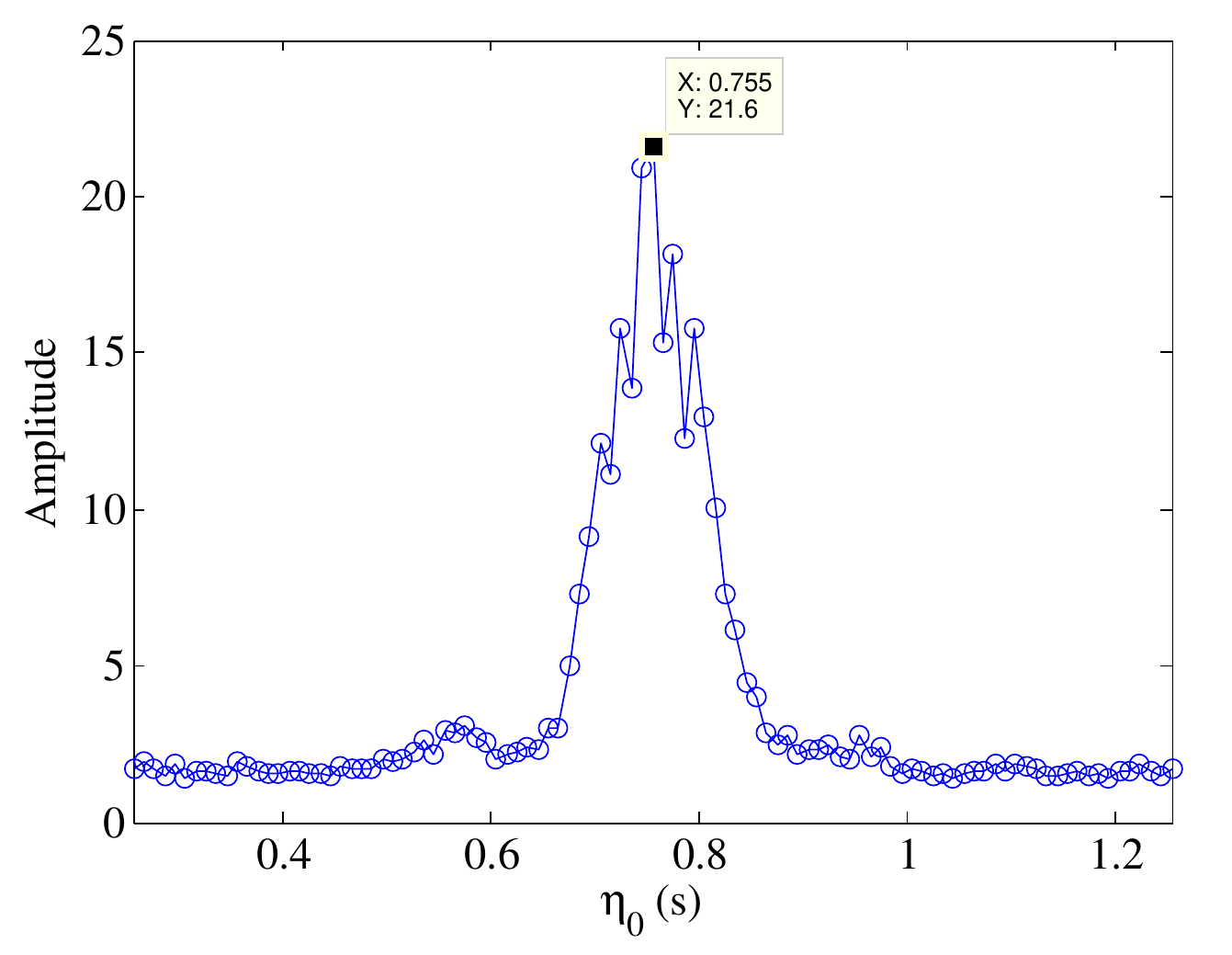}
\end{minipage}
}
\subfigure[]{
\begin{minipage}[b]{0.4\textwidth}\label{fig:different window:h}
\includegraphics[width=1.\textwidth,draft=false]{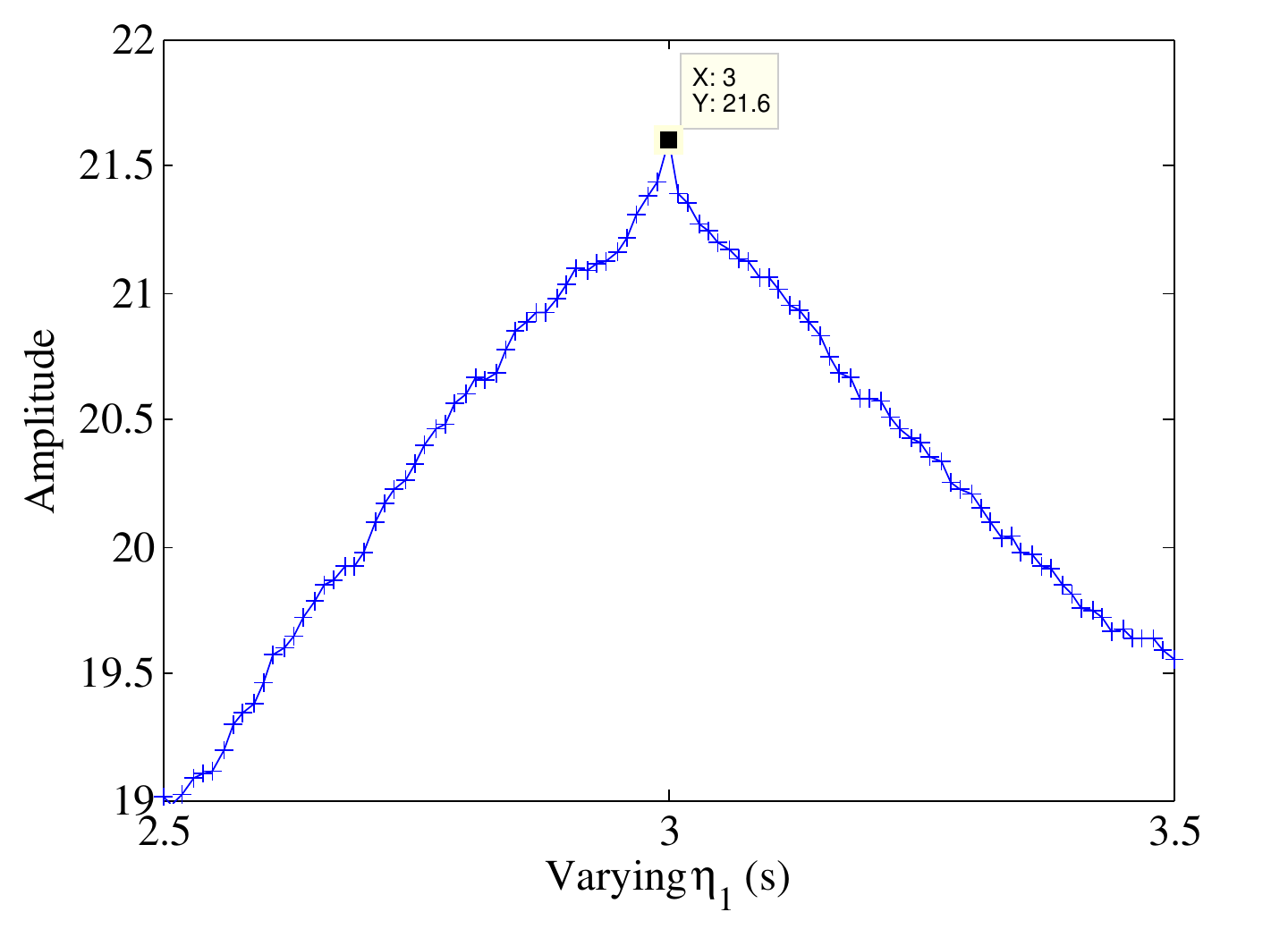}
\end{minipage}
}
\caption{WRFRFT response with respect to $\eta_0$ and $\eta_1$. (a) PC.
(b) $\eta_0=0.755s, \eta_1=3s$. (c) $\eta_0=0.15s, \eta_1=0.5s$. (d) $\eta_0=3.05s, \eta_1=3.5s$. (e) $\eta_0=0.505s, \eta_1=2.9s$. (f) $\eta_0=0.755s, \eta_1=3.4s$. (g) Integrated peak value curve for fixed $\eta_1$ but varying $\eta_0$. (h) Integrated peak value curve for
fixed $\eta_0$ but varying $\eta_1$.
 } \label{fig:WRFRFT with window firstly}
\end{figure*}

A simulation experiment (Table 1 shows the
radar parameters and Table 2 gives the target parameters) is given to show how the WRFRFT performs
with varying $\eta_0$ and $\eta_1$.
The result of the radar echo after pulse compression
is shown in Fig. \ref{fig:different window:a}.
The WRFRFT result when $\eta_0=0.755s$ and $\eta_1=3s$ (i.e.,
the WRFRFT's window function matches the beginning/ending
time of target signal)
is given in Fig. \ref{fig:different window:b} (slice of velocity and acceleration).
It is observed that the target signal is coherently
accumulated as a peak, which is corresponding to the
target's radial velocity and acceleration.
Fig. \ref{fig:different window:c} and Fig.
\ref{fig:different window:d} give respectively the WRFRFT results
for the cases that  $\eta_0=0.15s, \eta_1=0.5s$ and
$\eta_0=3.05s, \eta_1=3.5s$.
In these two cases (Fig. \ref{fig:different window:c}
and Fig. \ref{fig:different window:d}), only noise is extracted and
accumulated, resulting in unfocused results for WRFRFT.
Fig. \ref{fig:different window:e} gives the WRFRFT output when
$\eta_0=0.505s$ and $ \eta_1=2.9s$, where  only part of
the target signal is extracted in this case, and
thus the  peak value of Fig. \ref{fig:different window:e}
 is smaller than the peak value of
Fig. \ref{fig:different window:b}.
Fig. \ref{fig:different window:f} shows the WRFRFT output when  $\eta_0=0.755s, \eta_1=3.4s$,
where  the entire target signal is extracted
and coherently accumulated  in this case. However,
it is worth pointing out that compared with
Fig. \ref{fig:different window:b}, much more noise is extracted and accumulated
in Fig. \ref{fig:different window:f}. As a result,
the peak value of Fig. \ref{fig:different window:f} is
smaller than the peak value of Fig. \ref{fig:different window:b}.

In Fig. \ref{fig:different window:g}, we examine the WRFRFT outputs for
a fixed $\eta_1$
($\eta_1=3$) but varying $\eta_0$. In other words,
with the ending time of WRFRFT' window function $g(t)$ matches
the target signal's ending time,
the peak value reaches its maximum value when $\eta_0=0.755s$ (i.e.,
the beginning time of $g(t)$ equals to
the target signal's beginning time).
Similarly, Fig. \ref{fig:different window:h} shows the
integrated peak value  of WRFRFT for a fixed $\eta_0$ ($\eta_0=0.755s$) but varying $\eta_1$. That is to say,
the beginning time of $g(t)$ matches
the target signal's beginning time. We could see that
the peak value reaches its maximum value when $\eta_1=3s$ (i.e.,
the ending time of $g(t)$ equals to
the target signal's ending time).

Based on the results of Fig. \ref{fig:WRFRFT with window firstly}, it can be assured that
only when the beginning/ending time of WRFRFT's window function
equal to the target signal's beginning/ending time, then the integrated output of WRFRFT will reach its
maximum value.

\begin{table}[ht]
\begin{center}
\begin{tabular}{|c|c|}
\multicolumn{2}{c}{\textbf{Table \textbf{1}}
}\\
\multicolumn{2}{c}{Radar Parameters
}\\[5pt]
 \hline
Carrier frequency  &  6 GHz
\\
 \hline
Bandwidth  & 10 MHz
\\
 \hline
Sample frequency  & 50 MHz
\\
 \hline
Pulse repetition frequency  & 200 Hz
\\
 \hline
Pulse duration  & 10 us
\\
 \hline
Beginning time of target  & 0.755s
\\
 \hline
 SNR after PC  & 4 dB
\\
 \hline
\end{tabular}
\end{center}
\end{table}

\begin{table}[ht]
\begin{center}
\begin{tabular}{|c|c|}
\multicolumn{2}{c}{\textbf{Table \textbf{2}}
}\\
\multicolumn{2}{c}{Moving Target's Time Parameters and Motion Parameters
}\\[5pt]
 \hline
Initial range cell  &  287
\\
 \hline
Radial velocity ($\text{m}/\text{s}$)  & 90
\\
 \hline
Radial acceleration ($\text{m}/\text{s}^2$) & 26
\\
 \hline
Beginning time   & 0.755s
\\
 \hline
 Ending time   & 3s
\\
 \hline
\end{tabular}
\end{center}
\end{table}

\subsection{Procedure of the WRFRFT-based Method}\label{sec:procedure}

The main steps of the WRFRFT-based approach can be summarized
as follows:

Step 1:  Based on the relative prior information
of targets to be expected (such as moving status and
varieties) and the
radar parameters, the  searching
scope of initial range, velocity, acceleration and
beginning/ending time can be obtained (denoted respectively as
$[r_{\text{min}}, r_{\text{max}}]$, $[v_{\text{min}}, v_{\text{max}}]$, $[a_{\text{min}}, a_{\text{max}}]$, $[{\eta_0}_{\text{min}}, {\eta_0}_{\text{max}}]$
and  $[{\eta_1}_{\text{min}}, {\eta_1}_{\text{max}}]$).
In addition,
 the searching interval of beginning/ending time,
initial range, velocity and acceleration can
be set as \cite{ChenXL}:
\begin{equation}
\Delta \eta=PRT
\end{equation}
\begin{equation}
\Delta r=c/2B
\end{equation}
\begin{equation}
\Delta v=\lambda/2(T_1-T_0)
\end{equation}
\begin{equation}
\Delta a=\lambda/2(T_1-T_0)^2
\end{equation}
where $PRT$ is radar pulse repetition time.

Step 2: With the searching parameters
$(r_s,v_s,a_s,\eta_{0s},\eta_{1s})$, the moving trajectory
 to be searched and the window function $g(t)$ could be respectively
 expressed as:
\begin{equation}
r_s(t)=r_{0s}+v_s(t-\eta_{0s})+a_s(t-\eta_{0s})^2, t\in [\eta_{0s},\eta_{1s}]
\end{equation}
\begin{equation}
\begin{split}
g(t)=&\text{rect}\left[\frac{t-0.5(\eta_{1s}+\eta_{0s})}{\eta_{1s}-\eta_{0s}} \right]
\end{split}
\end{equation}
where $r_{0s}=r_{\text{min}}:\Delta r: r_{\text{max}}$,
$v_{s}=v_{\text{min}}:\Delta v: v_{\text{max}}$,
$a_{s}=a_{\text{min}}:\Delta a: a_{\text{max}}$,
$\eta_{0s}={\eta_{0}}_{\text{min}}:\Delta {\eta}: {\eta_{0}}_{\text{max}}$,
$\eta_{1s}={\eta_{1}}_{\text{min}}:\Delta {\eta}: {\eta_{1}}_{\text{max}}$.

Step 3: Intercept and extract the target signal from the compressed signal
based on the window function and searching motion trajectory, i.e.,
\begin{equation}
s_e(t)=g(t)s\left(2r_s(t)/c,t\right)
\end{equation}

Step 4: Apply the WRFRFT operation on the extracted signal.

Step 5: Go through all the searching parameters and obtain
the corresponding WRFRFT output $WR_{g(t)}(\alpha,u)$.

Step 6: Take the amplitude of WRFRFT output in step 5 as test
statistic, and compare with the adaptive threshold for a given
false alarm probability
\begin{equation}
|WR_{g(t)}(\alpha,u)|\test \gamma
\end{equation}
where $\gamma$ dentoes the detection threshold \cite{ChenXL},
which could be obtained via the reference
unit after WRFRFT.
If $|WR_{g(t)}(\alpha,u)|$ is larger than $\gamma$, target is
confirmed.
In addition, the target's time parameters and motion parameters can be
estimated via the peak location of $WR_{g(t)}(\alpha,u)$.

\begin{figure}[!htbp]
\begin{center}
\includegraphics[width=0.5 \textwidth,draft=false]{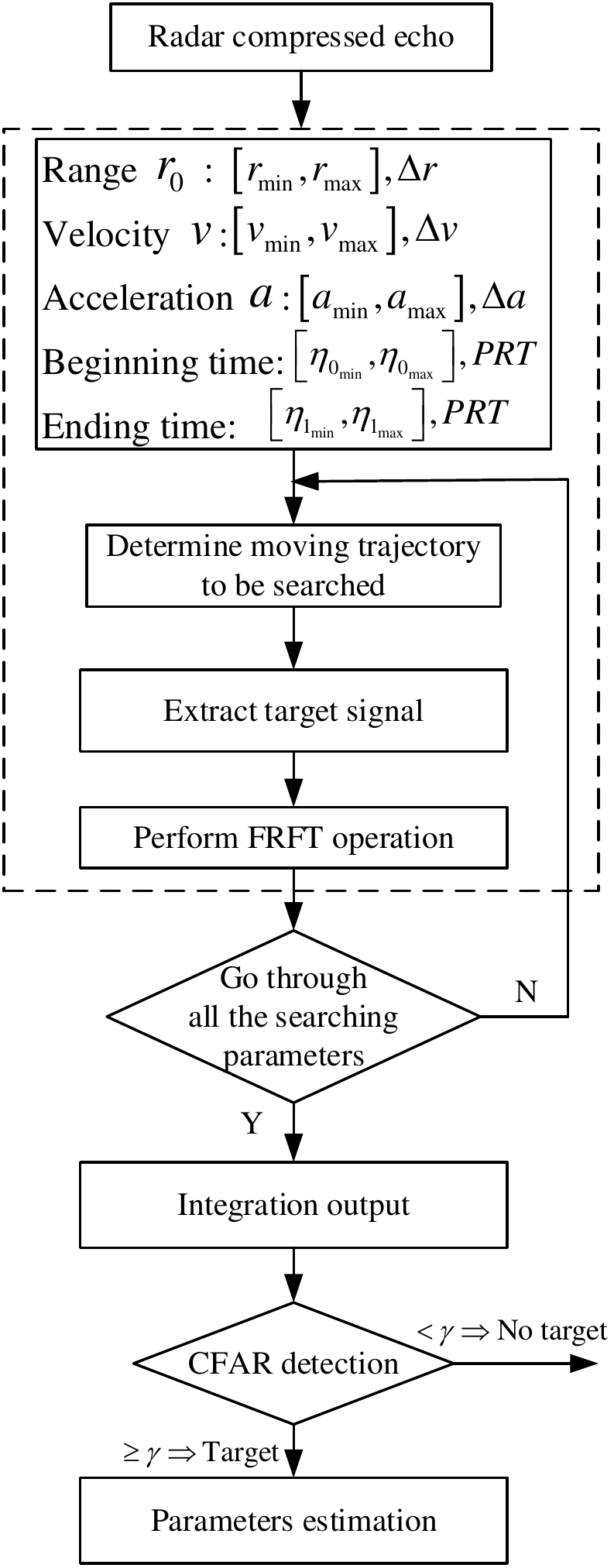}
\end{center}
\caption{Flowchart of the WRFRFT method.}\label{fig:flowchart}
\end{figure}

Fig. \ref{fig:flowchart} gives the flow chart of the WRFRFT-based method.

\subsection{Discussion on Computational Complexity}

It could be noticed that the  WRFRFT
mainly involves the
five dimensions searching (i.e., searching of beginning time, ending time,
range, velocity and acceleration) and the FRFT operation.
Note that the coherent integration of the target signal is
obtained when the searching time/motion parameters match with the target's
time/motion parameters and the fractional order of FRFT matches the target's
acceleration. Therefore, the following relationship could be
employed within the WRFRFT:
\begin{equation}
\alpha=\text{arccot} \left(\frac{-2a_sT_{\eta}}{\lambda f_s}\right)
\end{equation}
where $T_{\eta}=\eta_{1s}-\eta_{0s}+PRT$ and $f_s$ denotes the sample frequency.

According to the analysis in Section \ref{sec:procedure},
the searching numbers of
beginning time, ending time,  range,
 velocity and  acceleration are:
\begin{equation}
N_{\eta_0}=\text{round}\left(\frac{{\eta_0}_{\text{max}}-{\eta_0}_{\text{min}}}{\Delta \eta}\right)
\end{equation}
\begin{equation}
N_{\eta_1}=\text{round}\left(\frac{{\eta_1}_{\text{max}}-{\eta_1}_{\text{min}}}{\Delta \eta}\right)
\end{equation}
\begin{equation}
N_{r}=\text{round}\left(\frac{{r}_{\text{max}}-{r}_{\text{min}}}{\Delta r}\right)
\end{equation}
\begin{equation}
N_{v}=\text{round}\left(\frac{{v}_{\text{max}}-{v}_{\text{min}}}{\Delta v}\right)
\end{equation}
\begin{equation}
N_{a}=\text{round}\left(\frac{{a}_{\text{max}}-{a}_{\text{min}}}{\Delta a}\right)
\end{equation}
Then, the computational complexity of the WRFRFT-based approach is  $O(N_{\eta_0}N_{\eta_1}N_{r}N_{v}N_{a}N\text{log}_2N)$, where
$N$ is the pulse number.
The computational cost of RFRFT is in the order of $O(N_{r}N_{v}N_{a}N_{P}N\text{log}_2N)$ \cite{ChenXL},
where $N_P$ denotes the number of transform order of FRFT.
Hence, the computational burden of WRFRFT is bigger than that of RFRFT, since
the searching process of beginning/ending time.
It is necessary to study the fast implementation of WRFRFT method in the future.

\section{Experiments and Analysis}\label{sec:experiments}

The experiments with simulated data
(Section \ref{sec:a weak}, Section
\ref{sec:estimation},
Section \ref{sec:detection} and Section \ref{sec:multiple}) and
real data (Section \ref{sec:real}) are given to demonstrate the
effectiveness of the WRFRFT method,
where the radar parameters of Section \ref{sec:a weak}, Section
\ref{sec:estimation},
Section \ref{sec:detection} and Section \ref{sec:multiple}
are set as the same as those in Table 1.
In addition, the target's
time parameters and motion parameters
of Section \ref{sec:a weak}$-$Section
\ref{sec:estimation} are given in Table 2.
Several typical coherent detection algorithms
(RFRFT, GRFT, RFT) are used for comparison.

\subsection{WRFRFT  of Weak Target}\label{sec:a weak}

\begin{figure*}[!htbp]
\centering \subfigure[]{
\begin{minipage}[b]{0.4\textwidth}\label{fig:single firstly:a}
\includegraphics[width=1.\textwidth,draft=false]{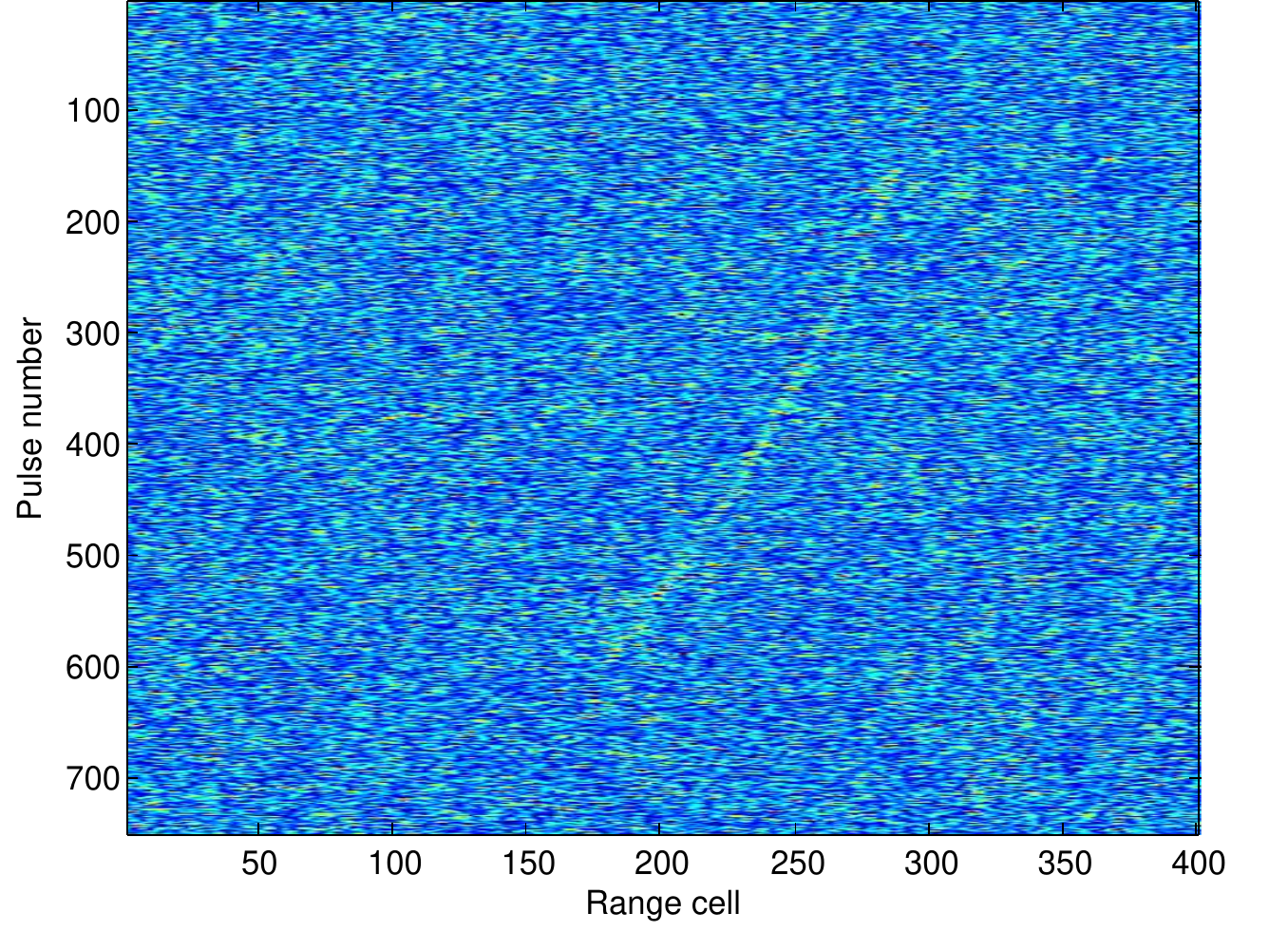}
\end{minipage}
} \subfigure[]{
\begin{minipage}[b]{0.4\textwidth}\label{fig:single firstly:b}
\includegraphics[width=1.\textwidth,draft=false]{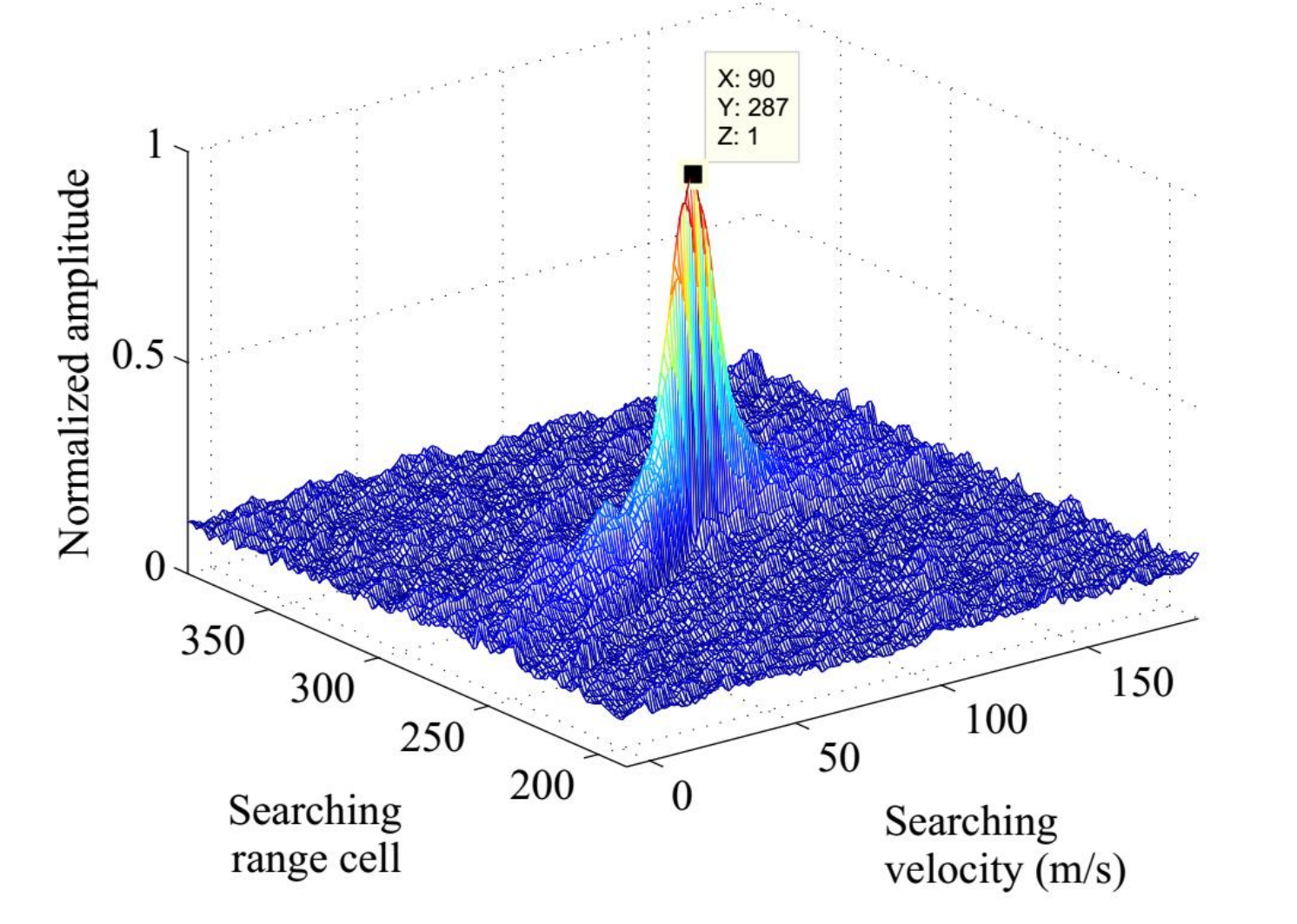}
\end{minipage}
} \subfigure[]{
\begin{minipage}[b]{0.4\textwidth}\label{fig:single firstly:c}
\includegraphics[width=1.\textwidth,draft=false]{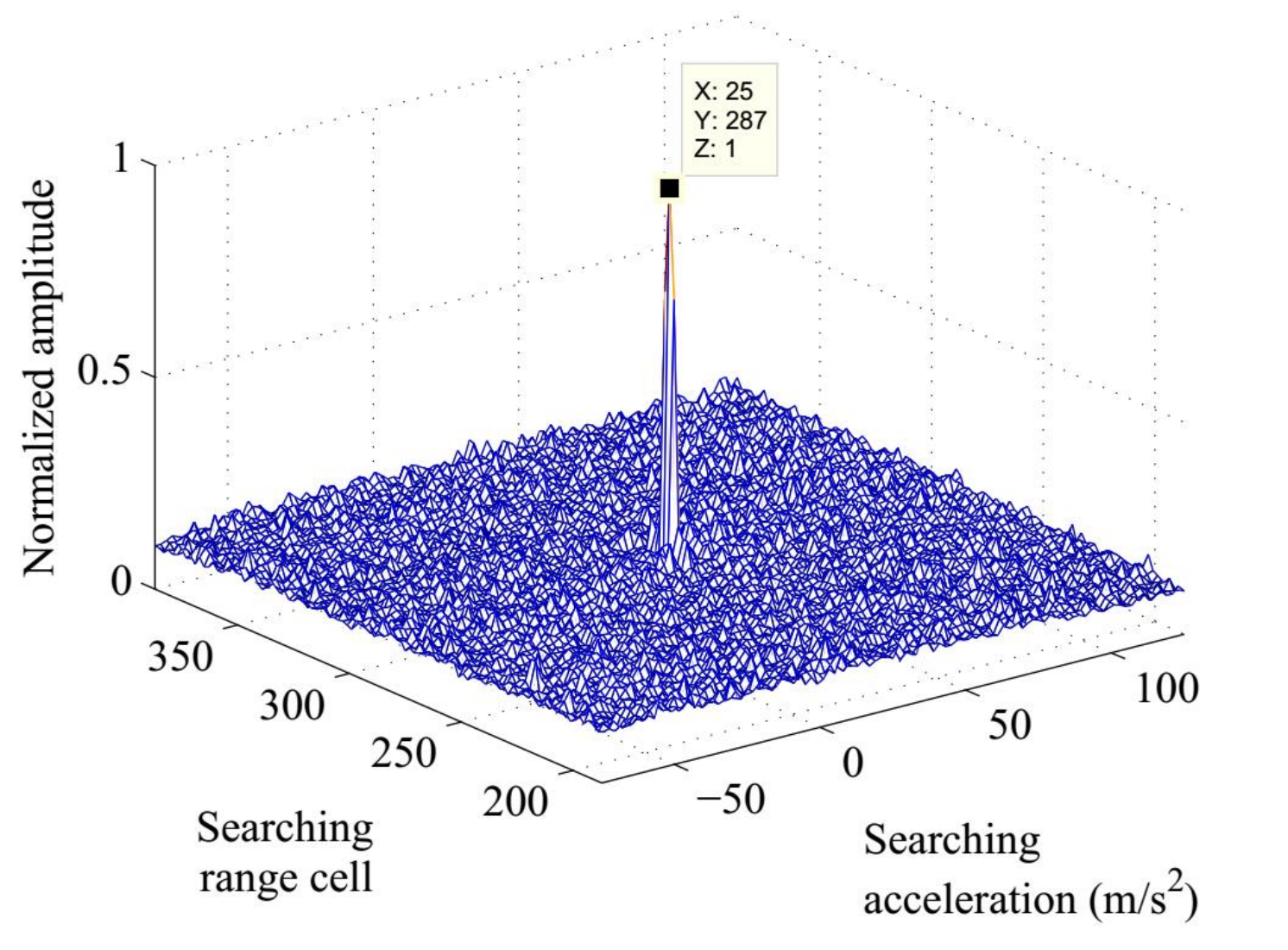}
\end{minipage}
}
\subfigure[]{
\begin{minipage}[b]{0.4\textwidth}\label{fig:single firstly:d}
\includegraphics[width=1.\textwidth,draft=false]{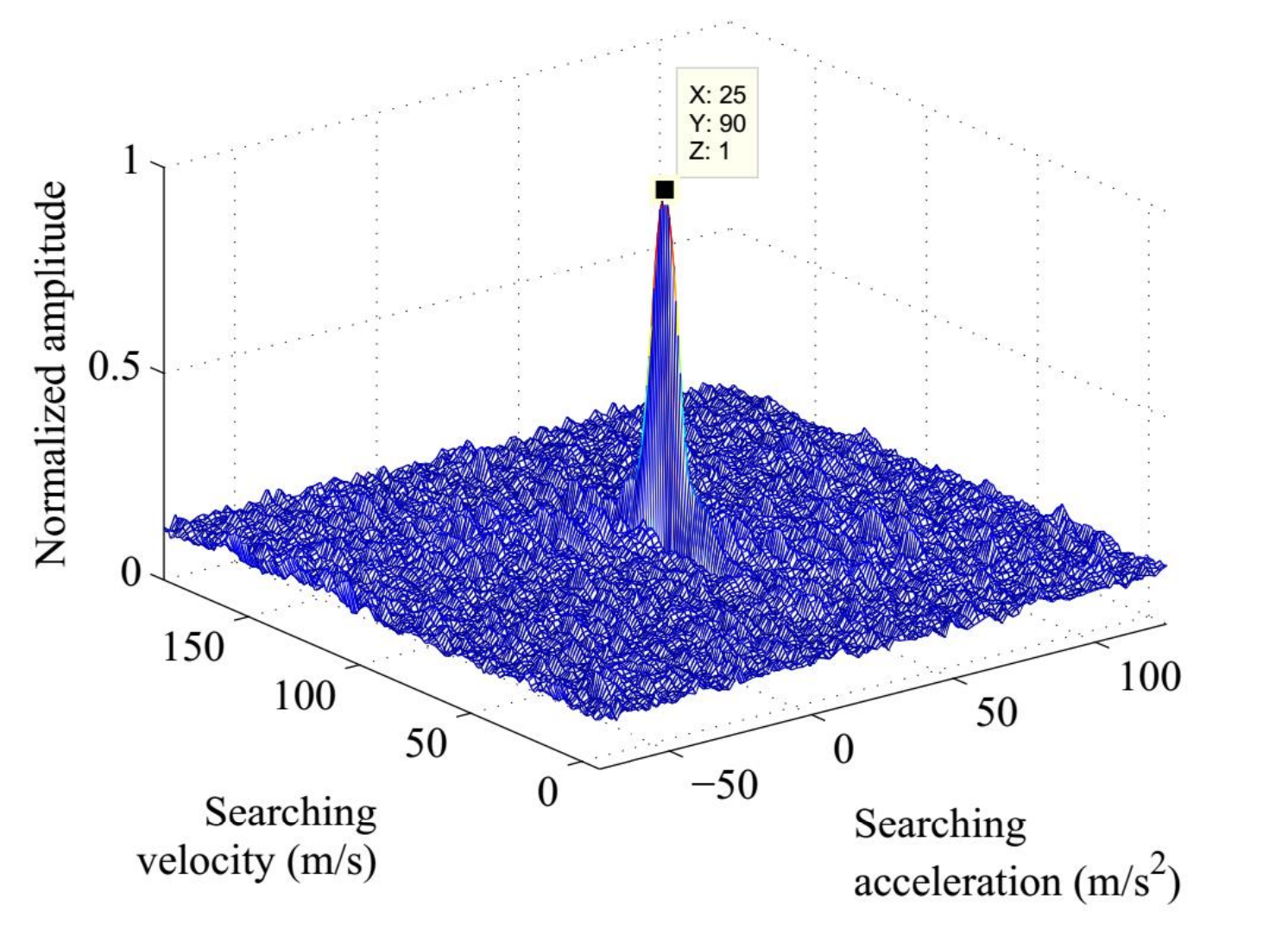}
\end{minipage}
}
\subfigure[]{
\begin{minipage}[b]{0.4\textwidth}\label{fig:single firstly:e}
\includegraphics[width=1.\textwidth,draft=false]{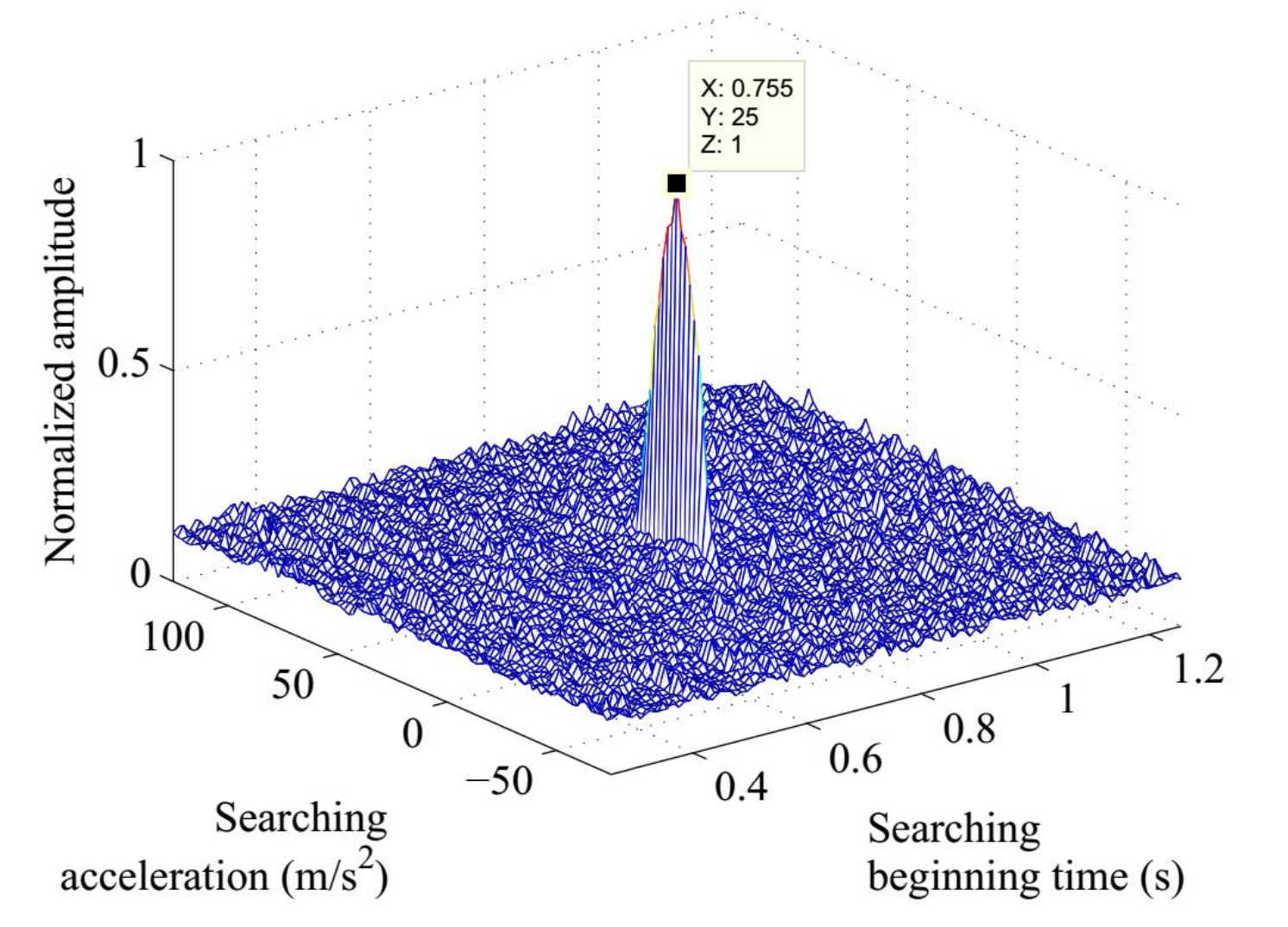}
\end{minipage}
}
\subfigure[]{
\begin{minipage}[b]{0.4\textwidth}\label{fig:single firstly:f}
\includegraphics[width=1.\textwidth,draft=false]{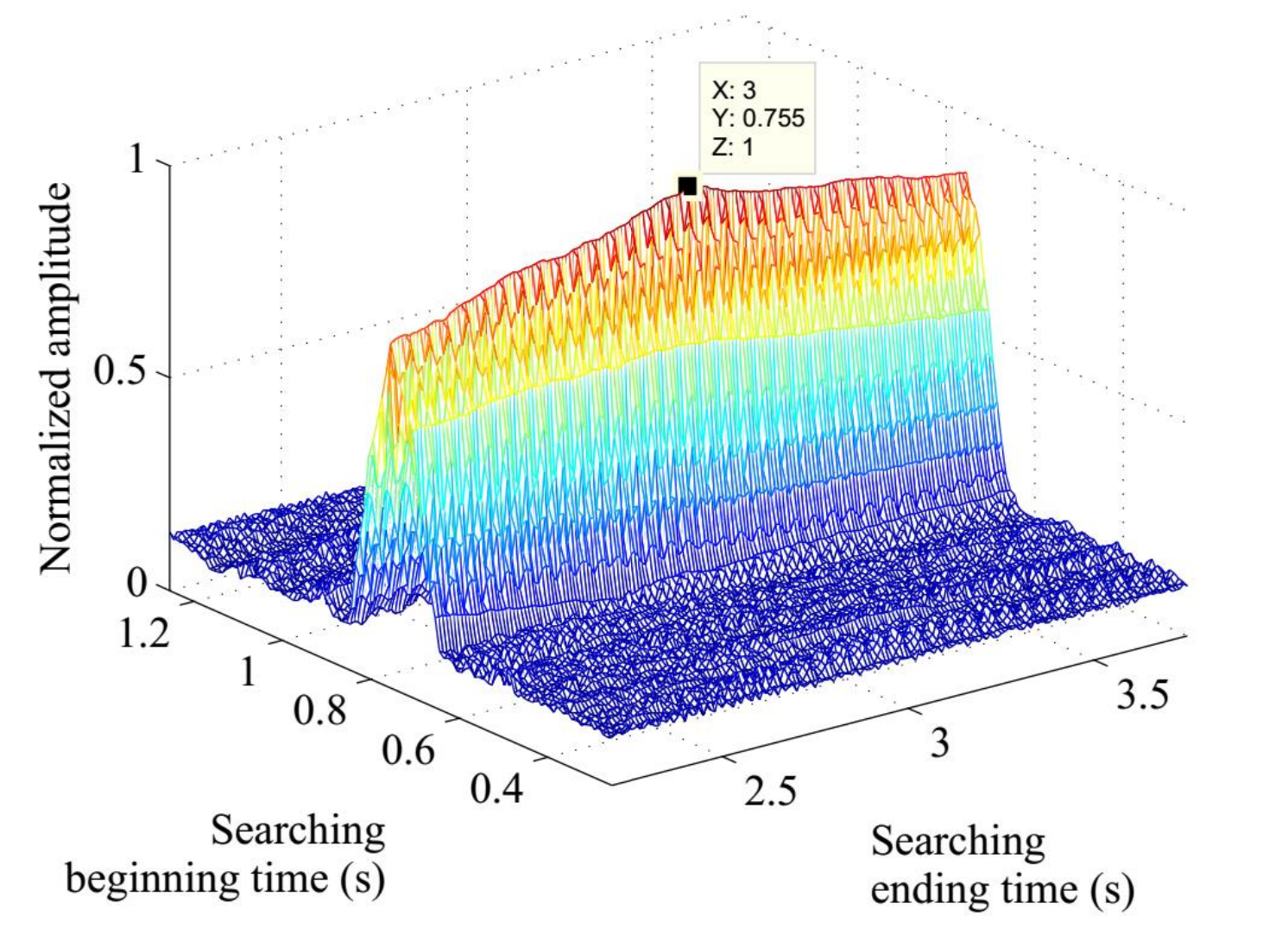}
\end{minipage}
}
\subfigure[]{
\begin{minipage}[b]{0.4\textwidth}\label{fig:single firstly:g}
\includegraphics[width=1.\textwidth,draft=false]{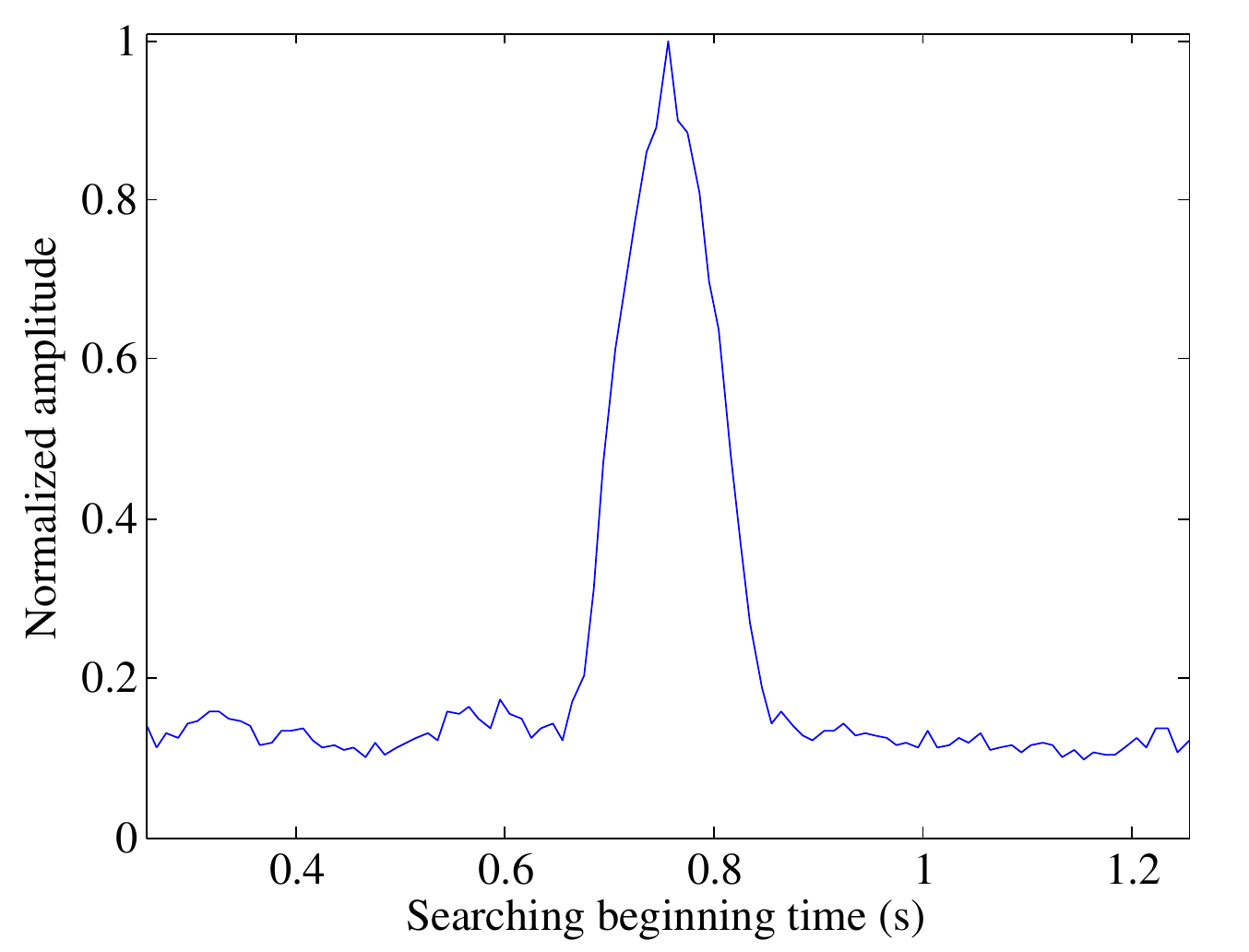}
\end{minipage}
}
\subfigure[]{
\begin{minipage}[b]{0.4\textwidth}\label{fig:single firstly:h}
\includegraphics[width=1.\textwidth,draft=false]{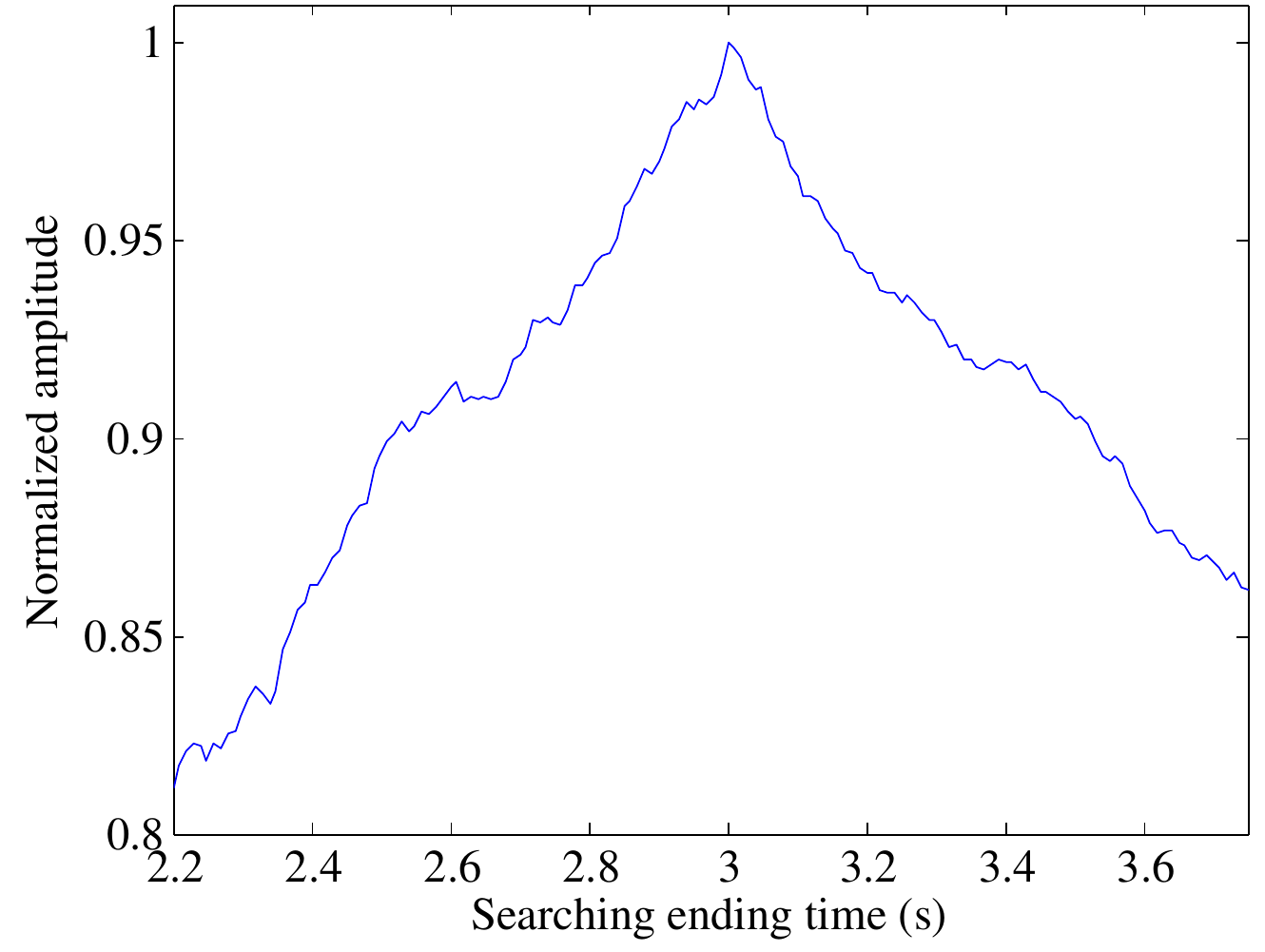}
\end{minipage}
}
\caption{WRFRFT for a weak target. (a) PC.
(b) Projection in  range cell-velocity domain.
(c) Projection in range cell-acceleration domain.
(d) Projection in  velocity-acceleration domain. (e)
Projection in acceleration-beginning time domain. (f)
Projection in beginning time-ending time domain. (g)
Beginning time response slice.
(h) Ending time response slice.
 } \label{fig:weak firstly}
\end{figure*}

In Fig. \ref{fig:weak firstly}, the WRFRFT's response for a weak target is given,
where the SNR after PC is 0 dB (as shown in Fig.
\ref{fig:single firstly:a}).
Fig. \ref{fig:single firstly:b}$-$Fig. \ref{fig:single firstly:f} give
respectively different projections of
the WRFRFT output.
More specifically, the projection  in range cell-velocity domain
is shown in Fig. \ref{fig:single firstly:b} and the projection in range cell-acceleration
domain is given in Fig. \ref{fig:single firstly:c}. Meanwhile, the projection in
velocity-acceleration space is shown in Fig. \ref{fig:single firstly:d} and
the projection in acceleration-beginning time domain
is given in Fig. \ref{fig:single firstly:e}. It can be noticed that
the peak locations of different projections (e.g., Fig. \ref{fig:single firstly:b}, Fig. \ref{fig:single firstly:c}, Fig.
\ref{fig:single firstly:d}) indicate
the corresponding parameters (e.g., range, velocity, acceleration,
beginning time) of target.
Also, the projection in beginning time-ending
time space is shown in Fig. \ref{fig:single firstly:f}, where Fig. \ref{fig:single firstly:g} and Fig. \ref{fig:single firstly:h}
give respectively the beginning time response slice and the ending time response
slice of Fig. \ref{fig:single firstly:f},
from which we could obtain the estimations of the beginning/ending time of
 target signal.

\begin{figure*}[!htbp]
\centering
\subfigure[]{
\begin{minipage}[b]{0.45\textwidth}\label{fig:single secondly:a}
\includegraphics[width=1.\textwidth,draft=false]{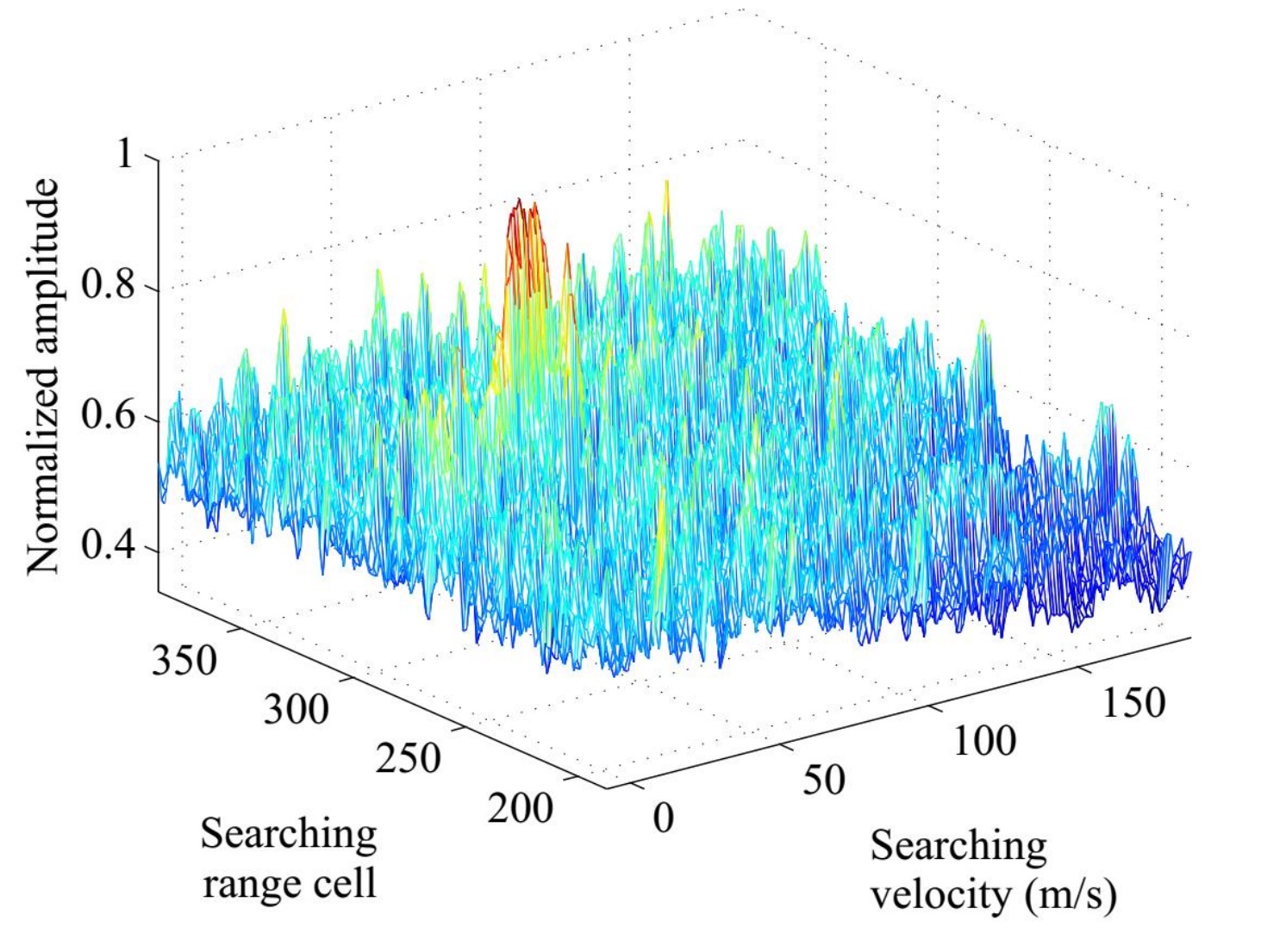}
\end{minipage}
} \subfigure[]{
\begin{minipage}[b]{0.45\textwidth}\label{fig:single secondly:b}
\includegraphics[width=1.\textwidth,draft=false]{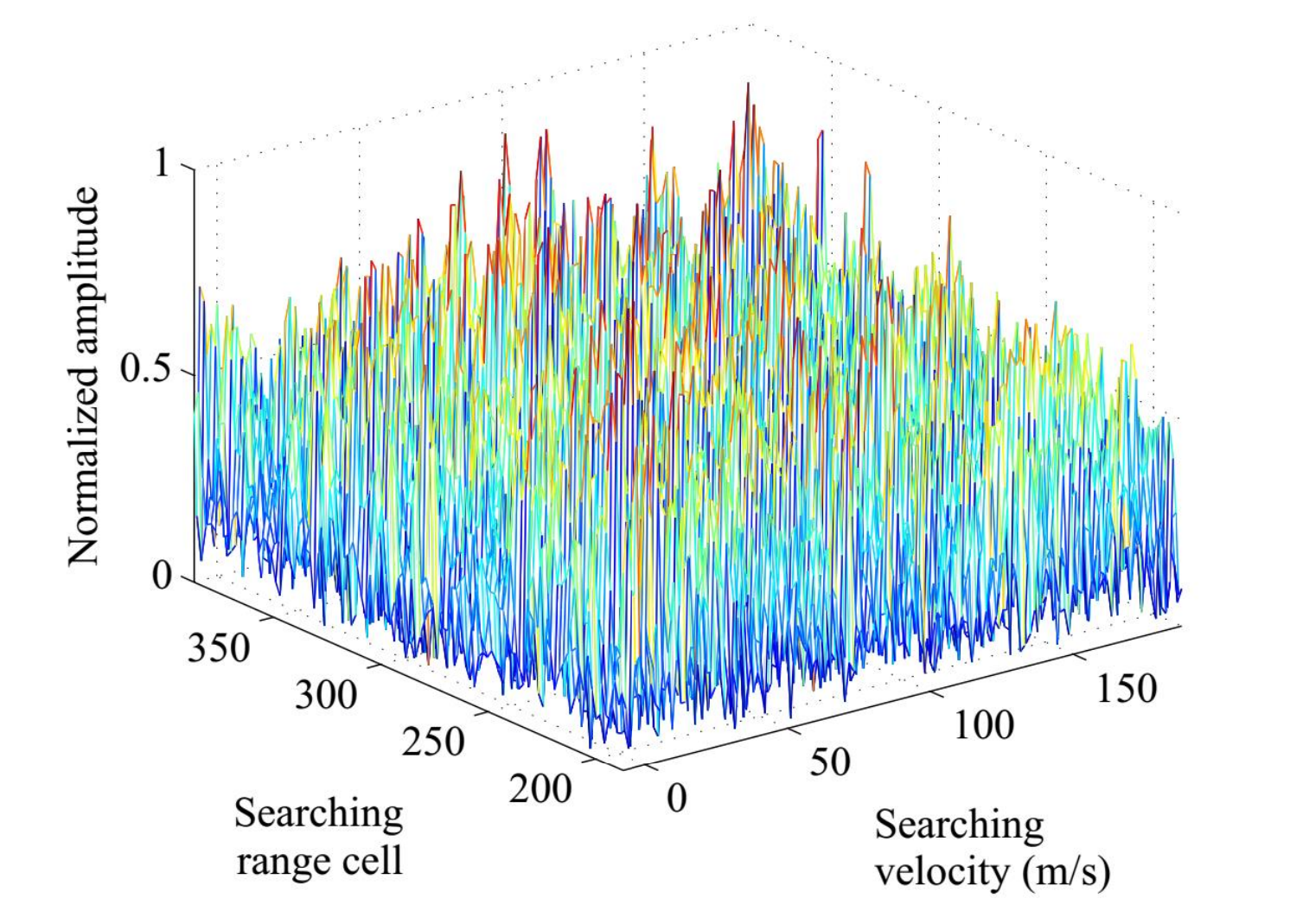}
\end{minipage}
}
\subfigure[]{
\begin{minipage}[b]{0.45\textwidth}\label{fig:single secondly:c}
\includegraphics[width=1.\textwidth,draft=false]{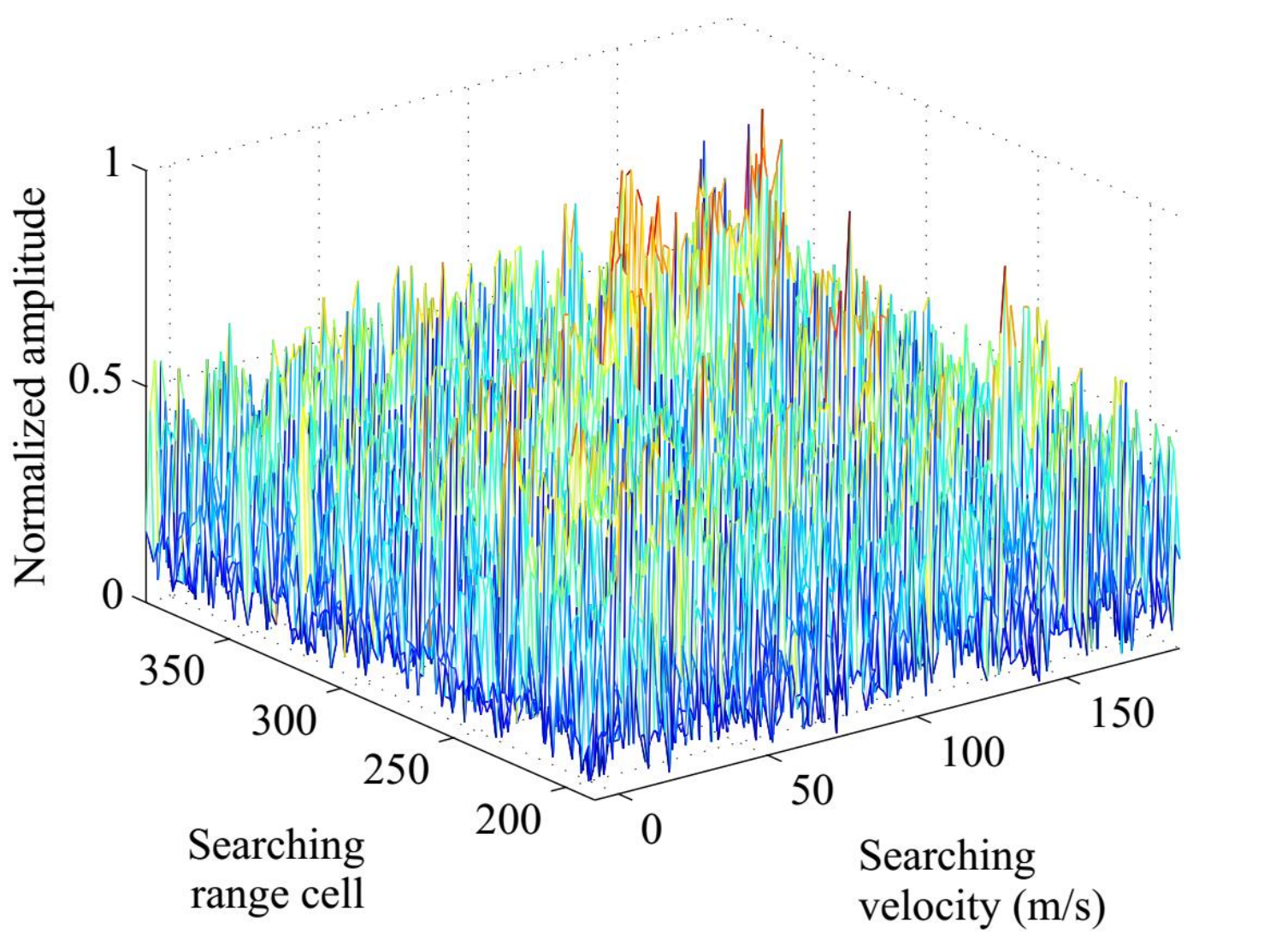}
\end{minipage}
}
\caption{Integration results of
(a) RFRFT. (b) GRFT. (c) RFT.
 } \label{fig:single secondly}
\end{figure*}

In order to comparison, the processing results of RFRFT, GRFT
and RFT
are  shown in Fig. \ref{fig:single secondly:a}$-$Fig. \ref{fig:single secondly:c}.
On the whole,
the outputs of these three algorithms are all defocused, due
to the mismatch among
the beginning/ending time of the target signal
and RFRFT, GRFT as well as the RFT.
Compared to the
integration results of GRFT and RFT
 (which are seriously defocused) the integration result of
RFRFT seems slightly better, but there is still no obvious
peak in Fig. \ref{fig:single secondly:a}.
More importantly, the relatively high peak position
of Fig. \ref{fig:single secondly:a}
does not appear at the position corresponding to the target's motion parameters,
which means that the target's motion parameters cannot be accurately estimated.

\subsection{Parameters Estimation Performance}\label{sec:estimation}

\begin{figure*}[!htbp]
\centering \subfigure[]{
\begin{minipage}[b]{0.45\textwidth}\label{fig:antinoise:a}
\includegraphics[width=1.\textwidth,draft=false]{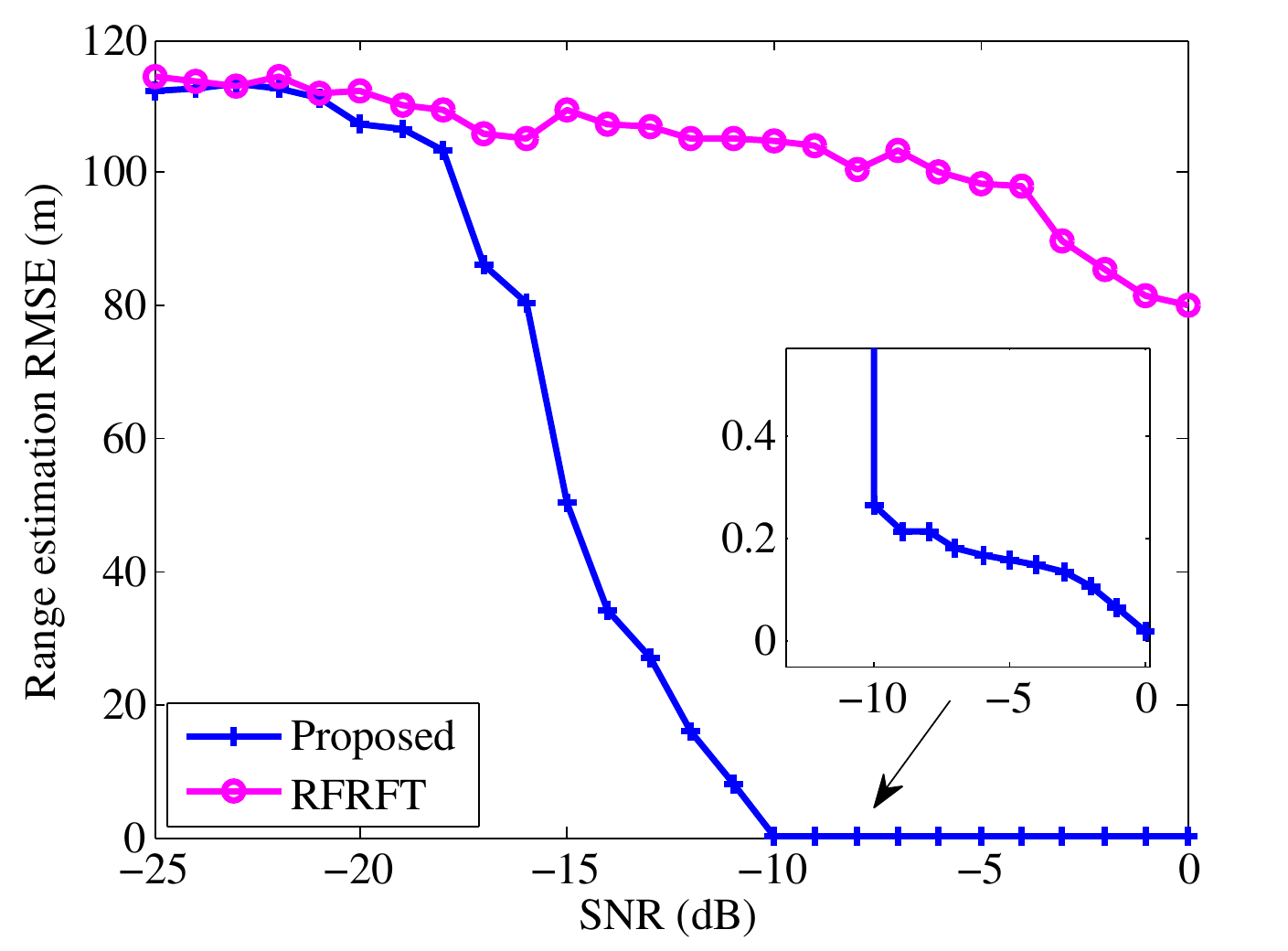}
\end{minipage}
} \subfigure[]{
\begin{minipage}[b]{0.45\textwidth}\label{fig:antinoise:b}
\includegraphics[width=1.\textwidth,draft=false]{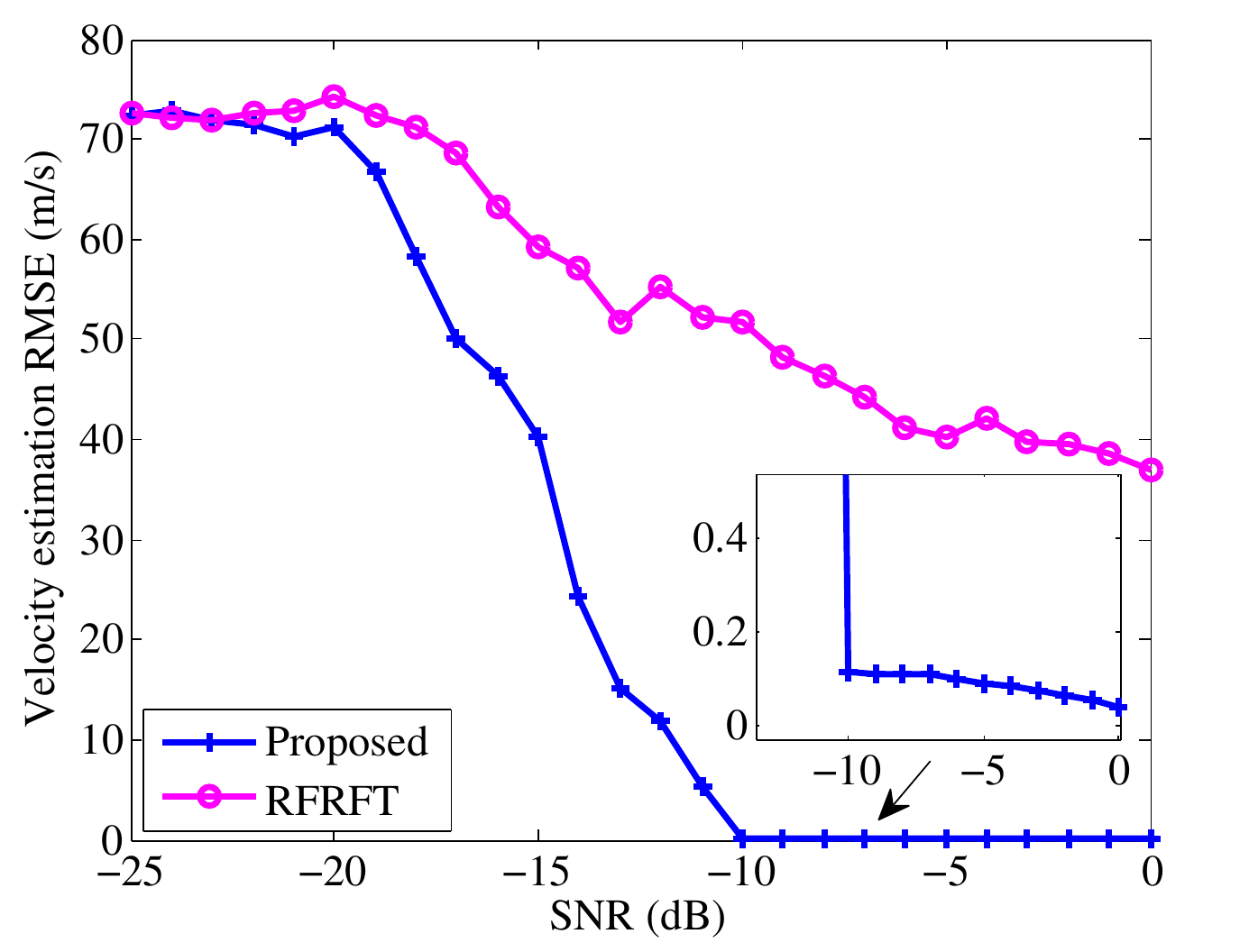}
\end{minipage}
}
\subfigure[]{
\begin{minipage}[b]{0.45\textwidth}\label{fig:antinoise:c}
\includegraphics[width=1.\textwidth,draft=false]{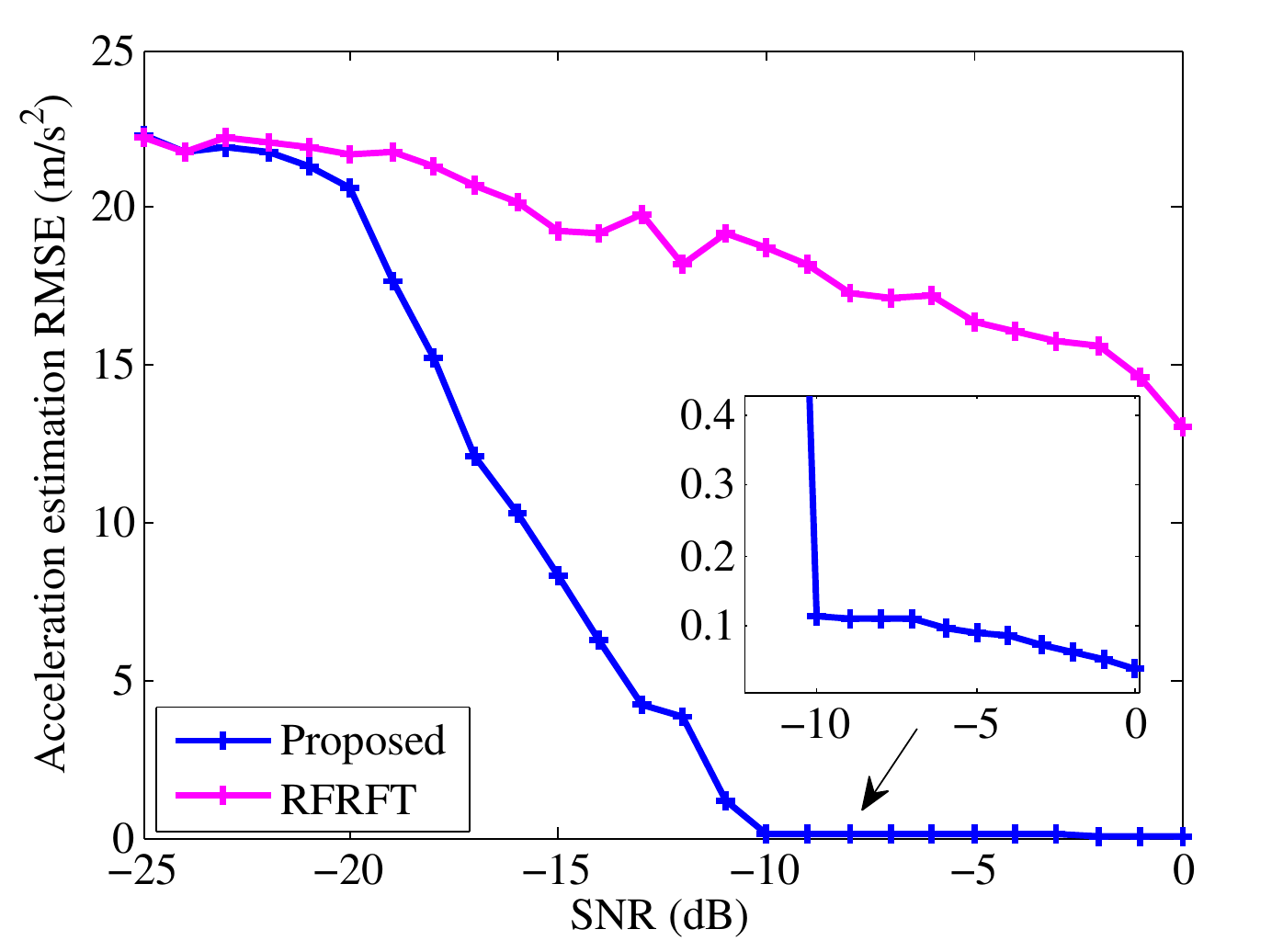}
\end{minipage}
}
\caption{RMSE of the estimated motion parameters. (a) Range.
(b) Velocity. (c) Acceleration.
 } \label{fig:RMSE}
\end{figure*}

\begin{figure*}[!htbp]
\centering \subfigure[]{
\begin{minipage}[b]{0.45\textwidth}\label{fig:time:a}
\includegraphics[width=1.\textwidth,draft=false]{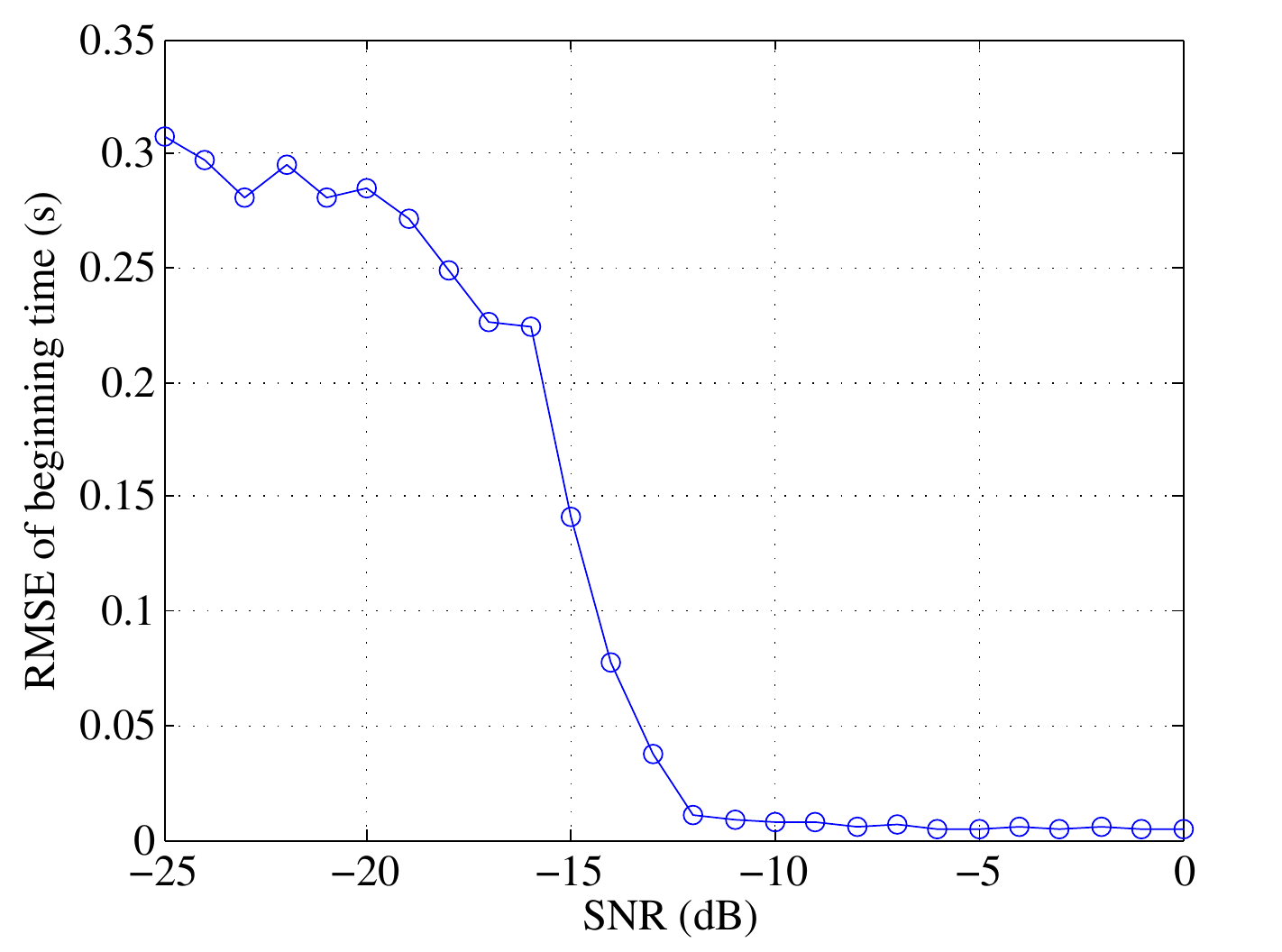}
\end{minipage}
} \subfigure[]{
\begin{minipage}[b]{0.45\textwidth}\label{fig:time:b}
\includegraphics[width=1.\textwidth,draft=false]{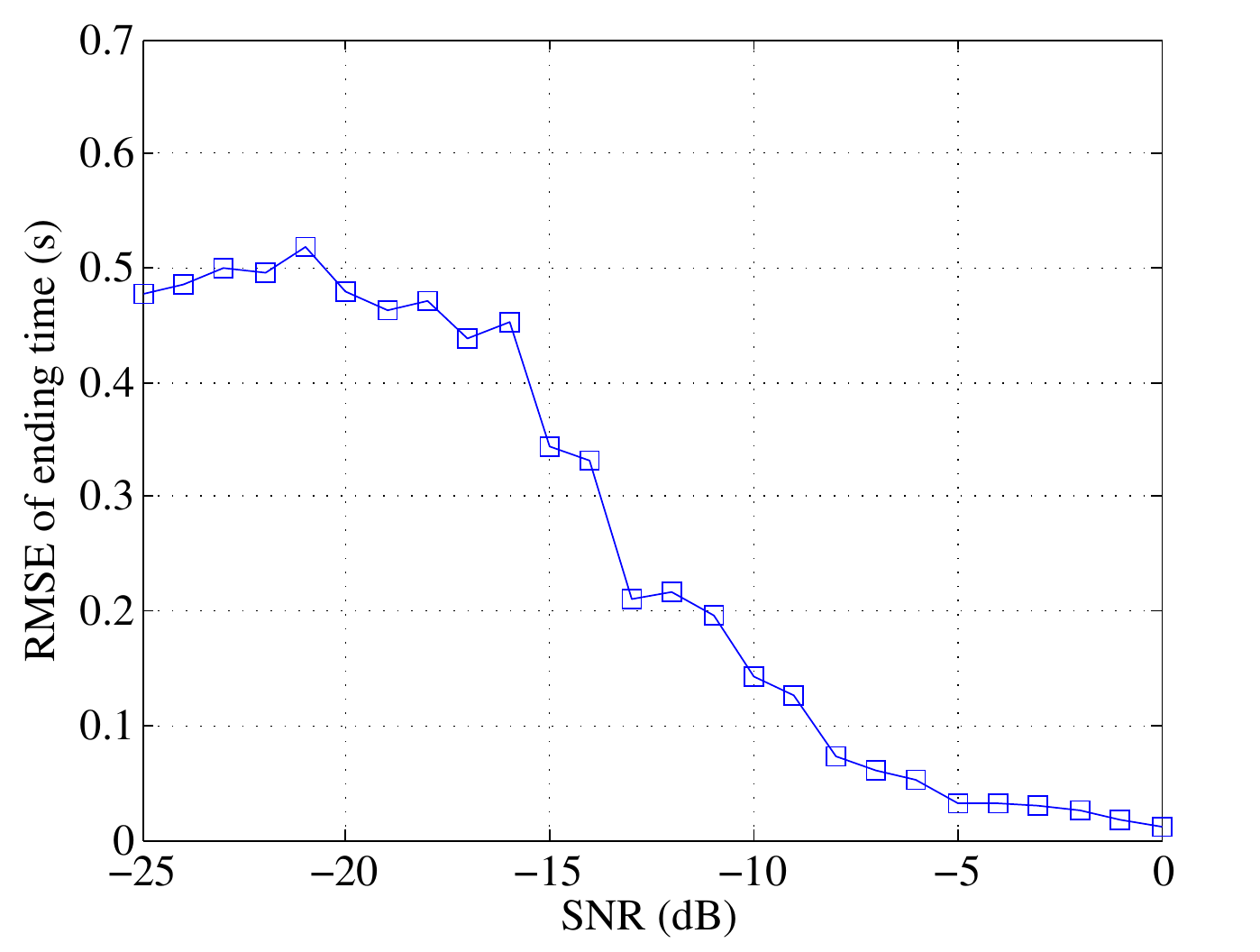}
\end{minipage}
}
\caption{RMSE of the estimated time parameters. (a) Beginning time (entry).
(b) Ending time (departure).
 } \label{fig:time RMSE}
\end{figure*}

The estimation performances
(i.e., root-mean-squared error (RMSE)) of the proposed WRFRFT algorithm for
target's motion parameters (range, velocity and acceleration) and time parameters (beginning/ending time) under different SNR levels are evaluated by
Monte Carlo  experiment.  Fig. \ref{fig:RMSE} gives the estimation performance curves
of target's motion parameters, where
200 times Monte Carlo trails are performed for each SNR.
From Fig. \ref{fig:RMSE}, we can notice that the WRFRFT
has a better estimation ability than RFRFT and
it
could obtain
good estimation performance when the SNR is larger than $-$10 dB.
Additionally, Fig. \ref{fig:time RMSE}
shows the RMSE curves of target's time parameters
where similar behaviors are observed.

\subsection{Detection Ability}\label{sec:detection}

\begin{figure}[!htbp]
\begin{center}
\includegraphics[width=0.75 \textwidth,draft=false]{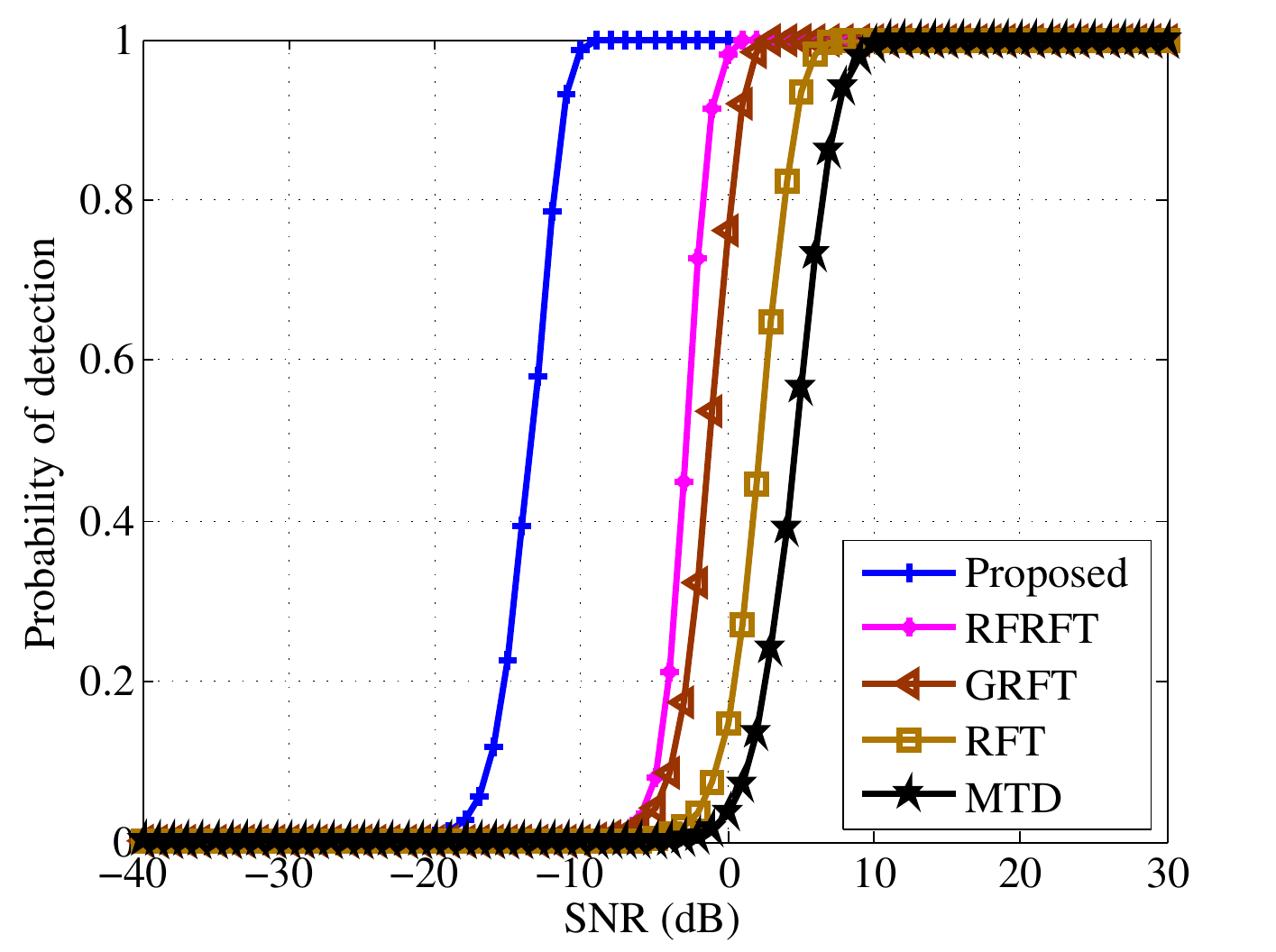}
\end{center}
\caption{Detection ability curves.}\label{fig:detection}
\end{figure}

The detection performances of the WRFRFT method, RFRFT, GRFT,
MTD and RFT under different
SNR  are shown in Fig. \ref{fig:detection}, where the false alarm rate is set as
$P_{f}=10^{-5}$. From Fig. \ref{fig:detection}, it could be noticed that the
proposed WRFRFT method could obtain better detection probability than RFRFT, GRFT and RFT
thanks to its ability of matching with the target's beginning time
and ending time as well as the correction of ARC and DS.
For example,   WRFRFT is able to detect a target with 0.8
 probability at SNR 10/11.5/15.6 dB lower than RFRFT/GRFT/RFT respectively.

\subsection{WRFRFT for Multiple Targets}\label{sec:multiple}

\begin{figure*}[!htbp]
\centering \subfigure[]{
\begin{minipage}[b]{0.45\textwidth}\label{fig:multi:a}
\includegraphics[width=1.\textwidth,draft=false]{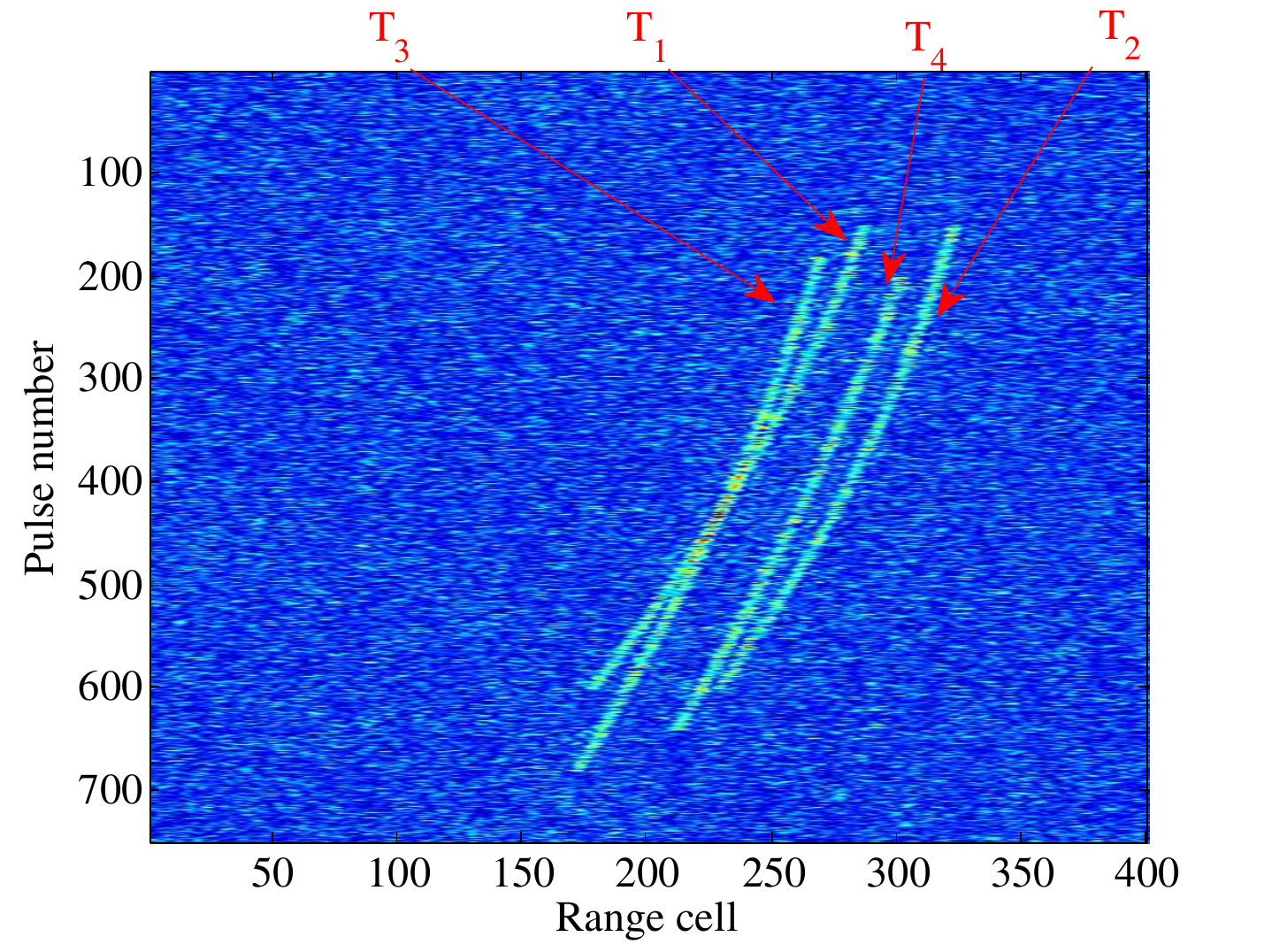}
\end{minipage}
} \subfigure[]{
\begin{minipage}[b]{0.45\textwidth}\label{fig:multi:b}
\includegraphics[width=1.\textwidth,draft=false]{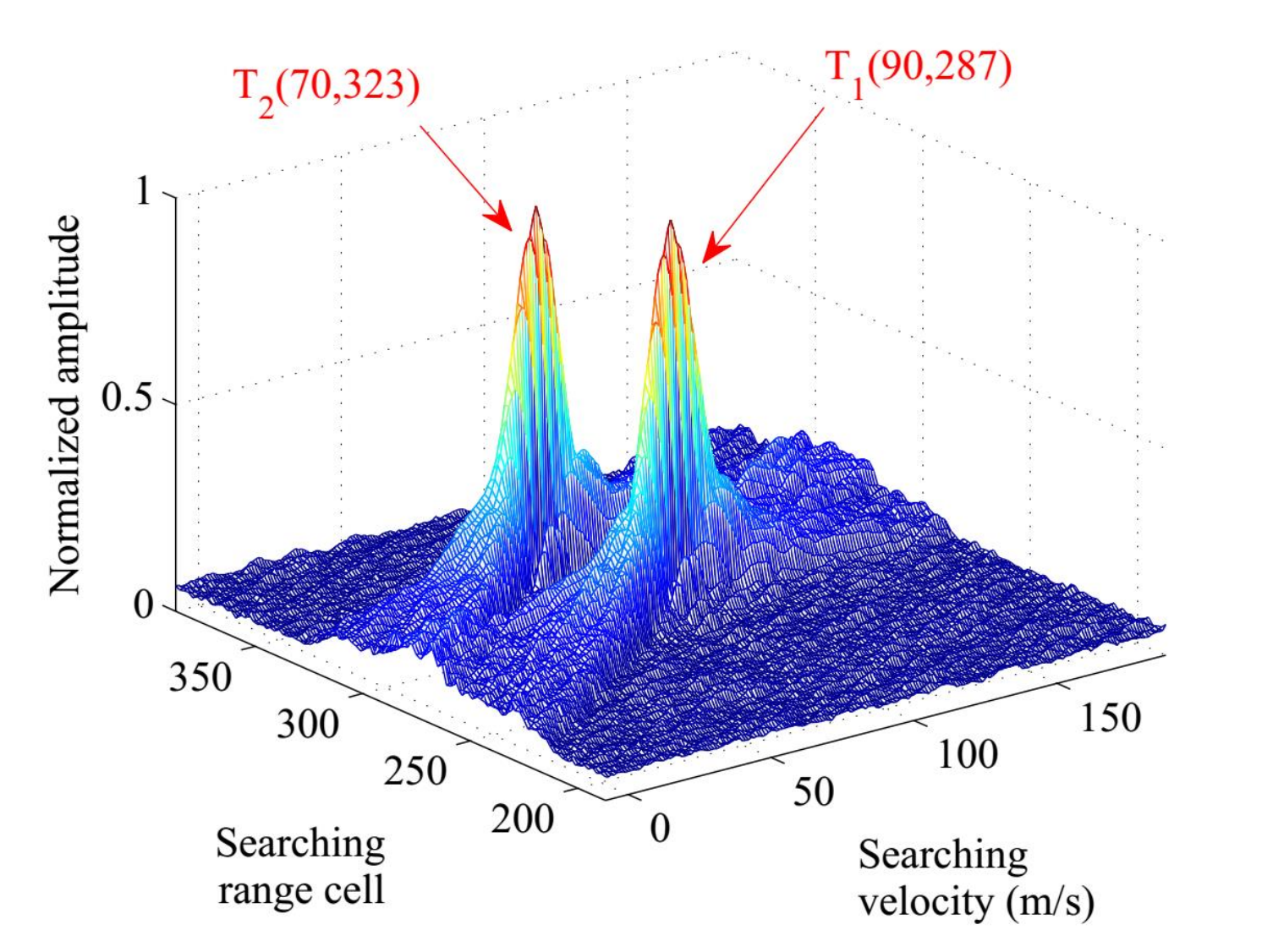}
\end{minipage}
}
\subfigure[]{
\begin{minipage}[b]{0.45\textwidth}\label{fig:multi:c}
\includegraphics[width=1.\textwidth,draft=false]{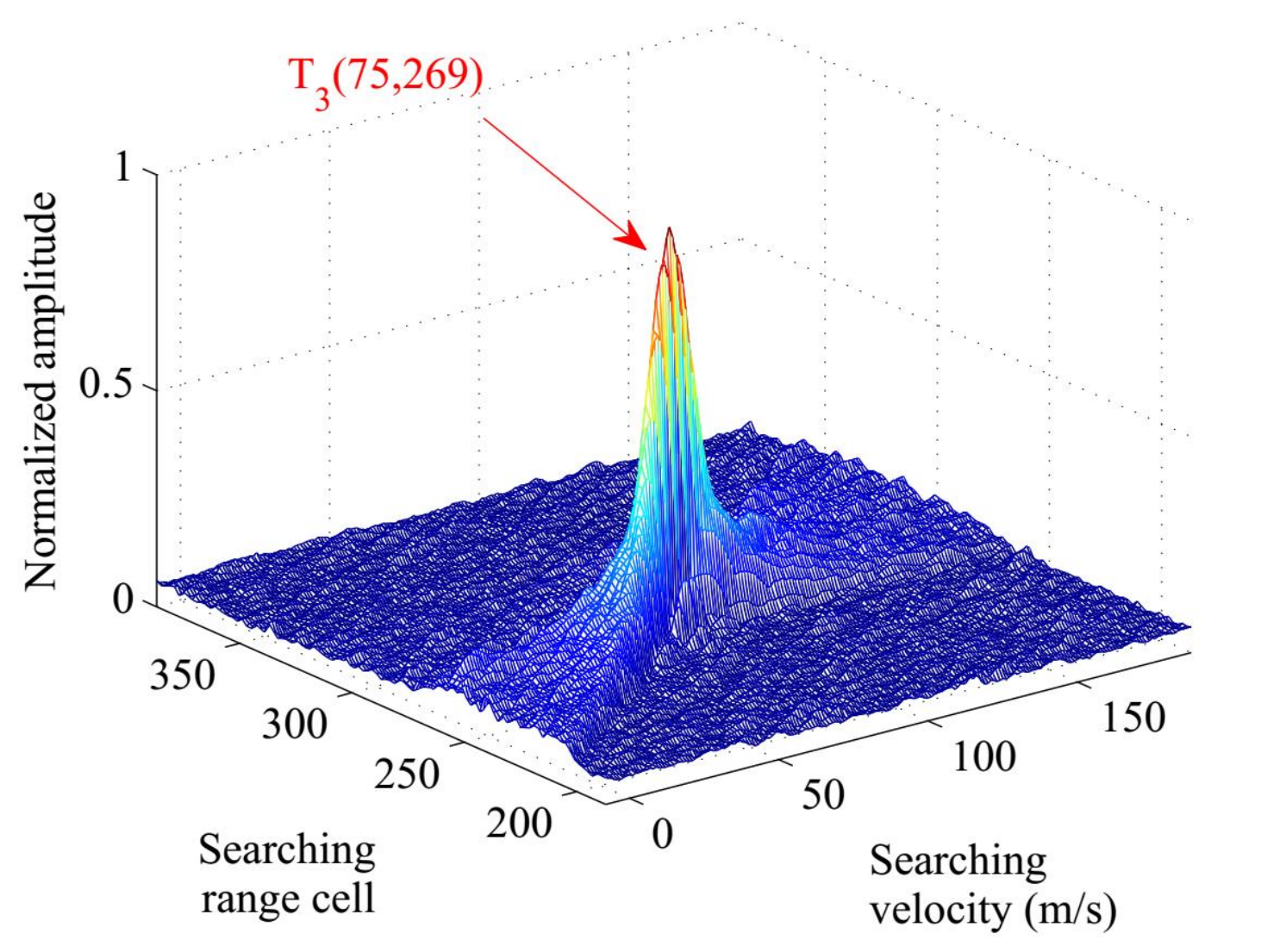}
\end{minipage}
}
\subfigure[]{
\begin{minipage}[b]{0.45\textwidth}\label{fig:multi:d}
\includegraphics[width=1.\textwidth,draft=false]{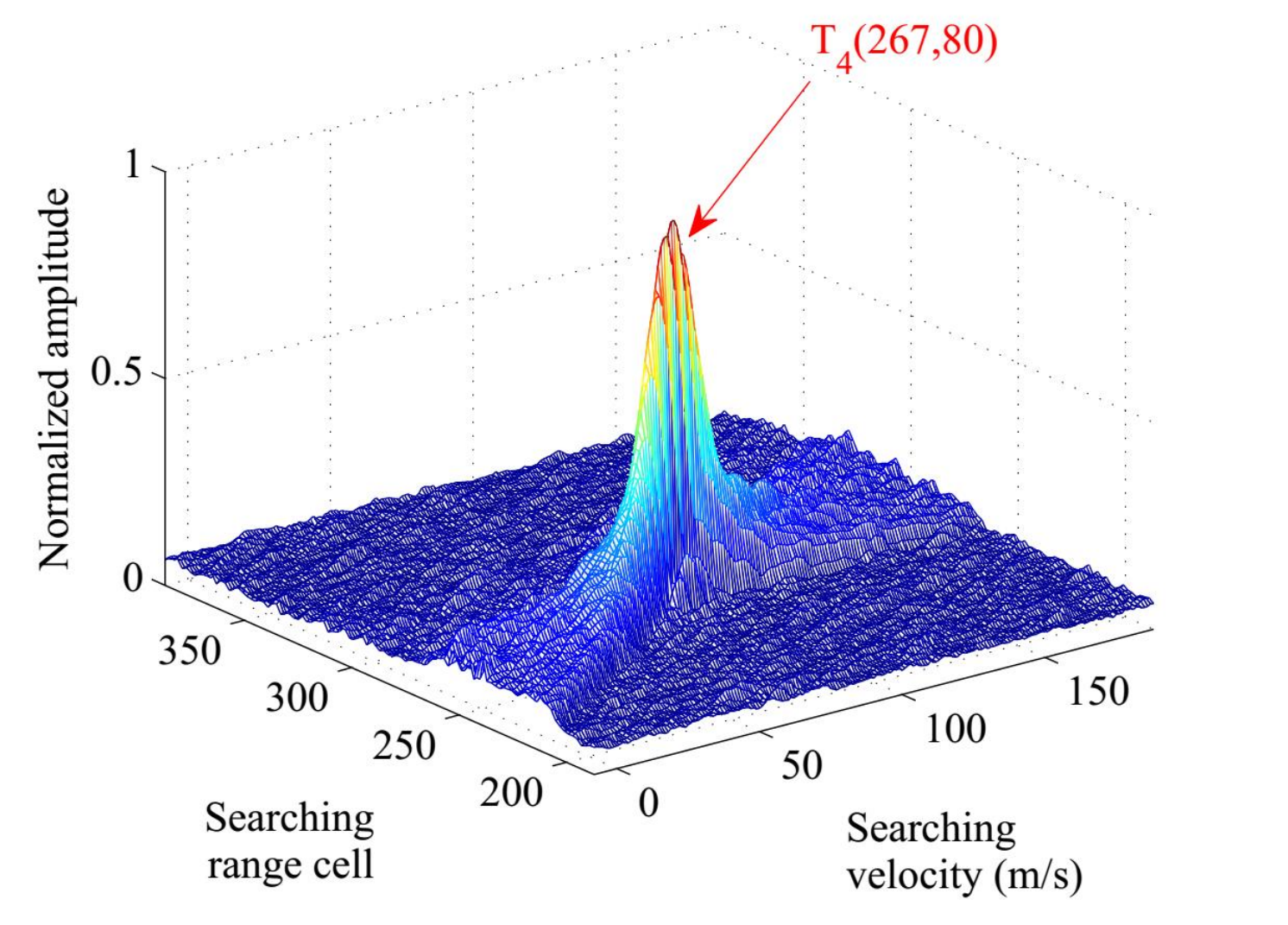}
\end{minipage}
}
\caption{Multiple targets scene. (a) Results after pulse compression.
(b)  The focusing result of $T_1$ and $T_2$
with $a_s=25\text{m}/\text{s}^2, \eta_{0s}=0.755\text{s}, \eta_{1s}=3\text{s}$. (c) The focusing result of $T_3$ with $a_s=17\text{m}/\text{s}^2, \eta_{0s}=0.905\text{s}, \eta_{1s}=3.4\text{s}$. (d)
The focusing result of $T_4$  with $a_s=13\text{m}/\text{s}^2, \eta_{0s}=1.005\text{s}, \eta_{1s}=3.2\text{s}$.} \label{fig:MULTIPLE}
\end{figure*}

Fig. \ref{fig:MULTIPLE} shows the simulation results of WRFRFT for multiple targets. Four moving targets
(denoted as $T_1$, $T_2$, $T_3$ and $T_4$) are considered and the
parameters are listed in Table 3.
The target motion trajectories after range compression
are shown in Fig. \ref{fig:multi:a}, where the four curved trajectories are observed.
Fig. \ref{fig:multi:b}$-$Fig. \ref{fig:multi:d} give respectively different slices of WRFRFT output.

More specifically, Fig. \ref{fig:multi:b} shows the WRFRFT result with $a_s=25\text{m}/\text{s}^2$, $\eta_{0s}=0.755\text{s}$, $\eta_{1s}=3\text{s}$. Note that the
searching values of acceleration, beginning/ending time are match with
the corresponding parameters of $T_1$ and $T_2$. Hence, the signal energy of
$T_1$ and $T_2$ are coherently accumulated and the targets are are well focused
 as seen from the two peaks in this
slice (Fig. \ref{fig:multi:b}). However, because of the  searching values of acceleration, beginning time and ending time
in this slice are not matched with $T_3$ and $T_4$ and thus the signal energy of
$T_3$ and $T_4$ can not be coherently integrated
in this slice.

Fig. \ref{fig:multi:c} shows the
WRFRFT result with $a_s=17\text{m}/\text{s}^2$, $\eta_{0s}=0.905\text{s}$, $\eta_{1s}=3.4\text{s}$,
which is matched with the acceleration and beginning/ending time of $T_3$. Correspondingly,
the target signal of $T_3$ is coherently focused as a peak in this slice (Fig. \ref{fig:multi:c}).
Moreover,
Fig. \ref{fig:multi:d} shows the
WRFRFT result with $a_s=13\text{m}/\text{s}^2$, $\eta_{0s}=1.005\text{s}$, $\eta_{1s}=3.2\text{s}$.
In this slice (Fig. \ref{fig:multi:d}),
the searching values of acceleration and beginning/ending time
match with the parameters of $T_4$ and thus we could notice that there is a peak formed,
which is corresponding to $T_4$.

\begin{table}[ht]
\begin{center}
\begin{tabular}{|c|c|c|c|c|}
\multicolumn{5}{c}{\textbf{Table \textbf{3}}
}\\
\multicolumn{5}{c}{Parameters of the Four Moving Targets
}\\[5pt]
 \hline
Parameter  &  $T_1$ & $T_2$ & $T_3$ & $T_4$
\\
 \hline
Initial range cell  & 287  & 323   & 269 & 305
\\
 \hline
Velocity (m/s) & 90  & 70  & 75 &95
\\
 \hline
Acceleration $\text{m}/{s}^2$ & 25  &  25   & 17 &13
\\
 \hline
Beginning time (s)  & 0.705  & 0.705  & 0.905 & 1.005
\\
 \hline
Ending time (s) & 3 & 3   & 3.4 & 3.2
\\
 \hline
 SNR after PC  & 6dB & 6dB   & 6dB & 6dB
\\
 \hline
\end{tabular}
\end{center}
\end{table}

\subsection{Real Data Results}\label{sec:real}

To verify the effectiveness of the WRFRFT method, an
evaluation is made  using real collected data. The data
set is the target signal of an unmanned aerial vehicle (UAV). Its maximum speed is
60 $\text{m}/\text{s}$ and its maximum acceleration is 10 $\text{m}/\text{s}^2$.
The detection system was a LFM pulse radar, the detailed parameters
are listed in Table 4. During the experiment,
the target did not appear in the radar detection beam
for the early stage of radar startup,
so there was only noise and clutter, but no target echo signal.
Then, the UAV flew into the radar detection area,
and the radar received the echo signal of the UAV.
The UAV flew out of the radar detection area
a few seconds later, but the radar was still transmitting and receiving signals
after the UAV's departure.

\begin{figure*}[!htbp]
\centering \subfigure[]{
\begin{minipage}[b]{0.45\textwidth}\label{fig:real:a}
\includegraphics[width=1.\textwidth,draft=false]{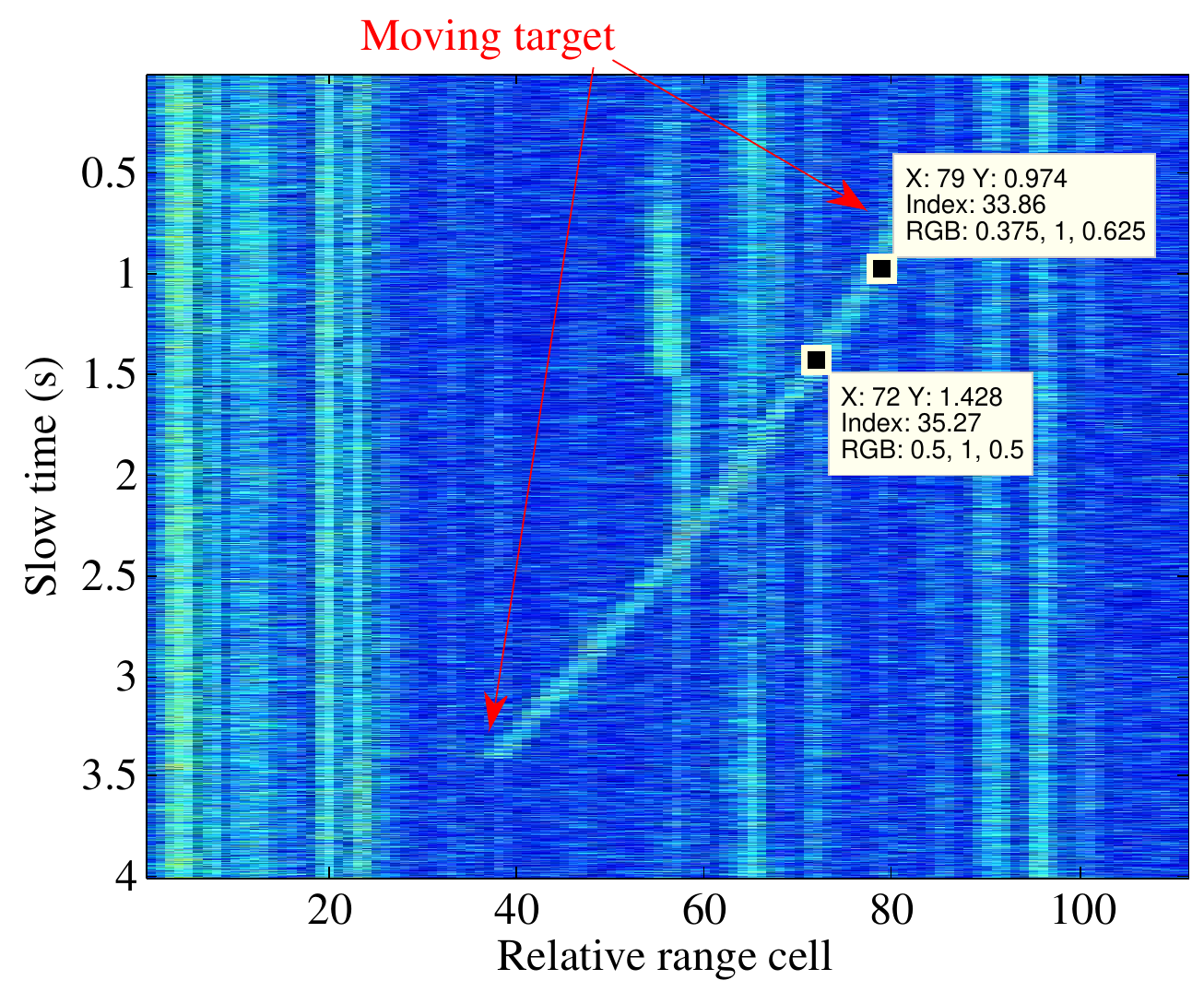}
\end{minipage}
} \subfigure[]{
\begin{minipage}[b]{0.45\textwidth}\label{fig:real:b}
\includegraphics[width=1.\textwidth,draft=false]{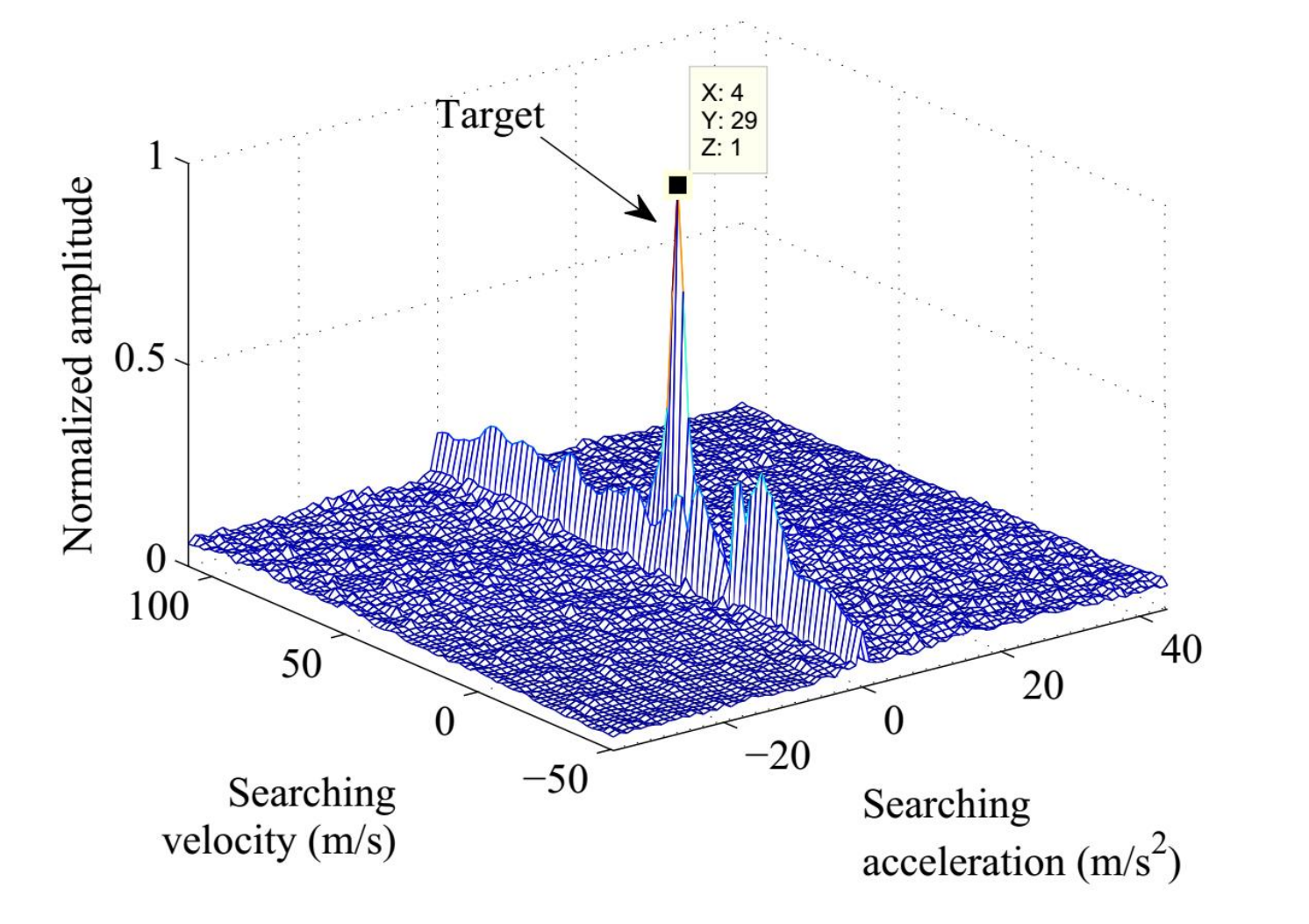}
\end{minipage}
}
\subfigure[]{
\begin{minipage}[b]{0.45\textwidth}\label{fig:real:c}
\includegraphics[width=1.\textwidth,draft=false]{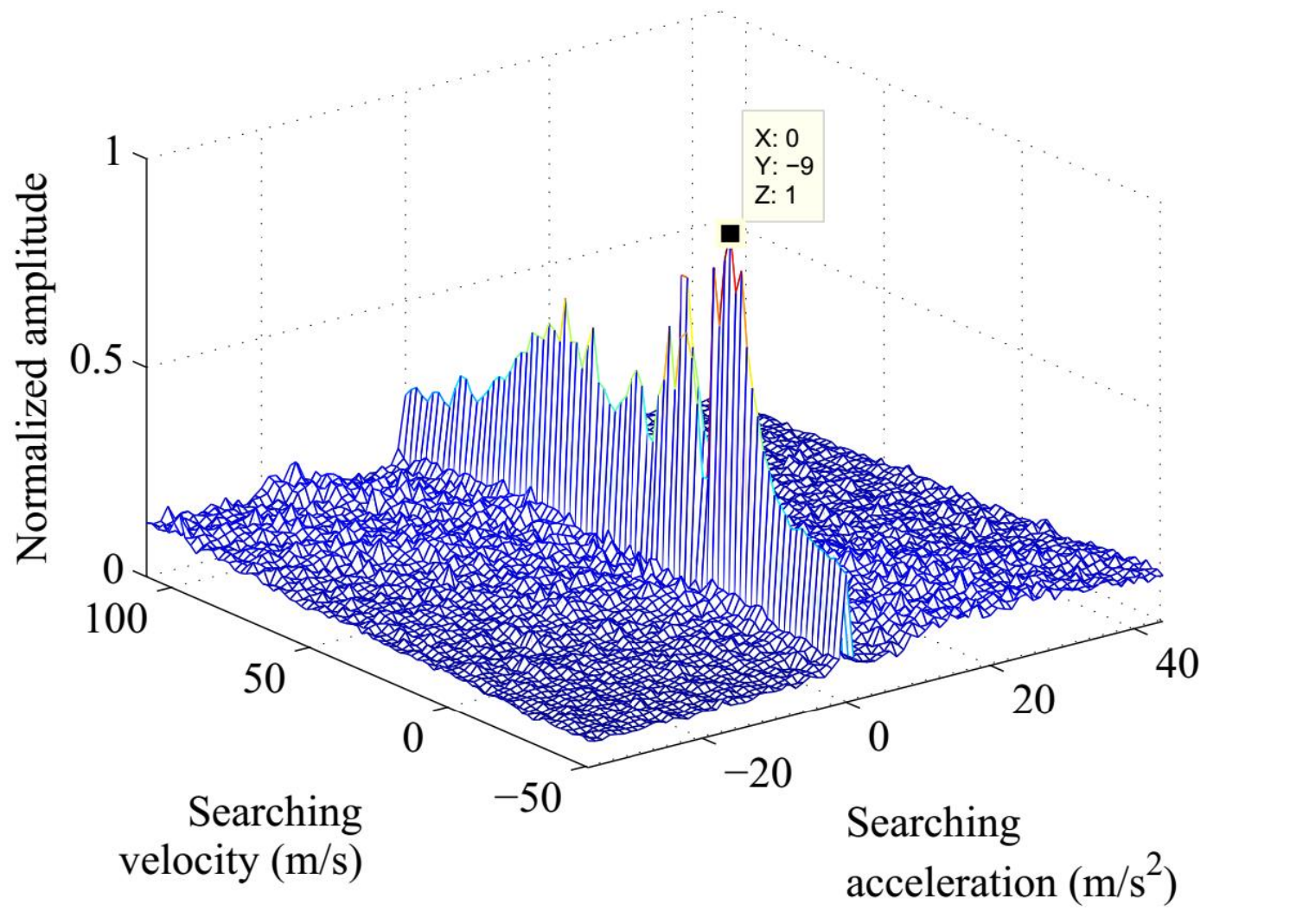}
\end{minipage}
}
\subfigure[]{
\begin{minipage}[b]{0.45\textwidth}\label{fig:real:d}
\includegraphics[width=1.\textwidth,draft=false]{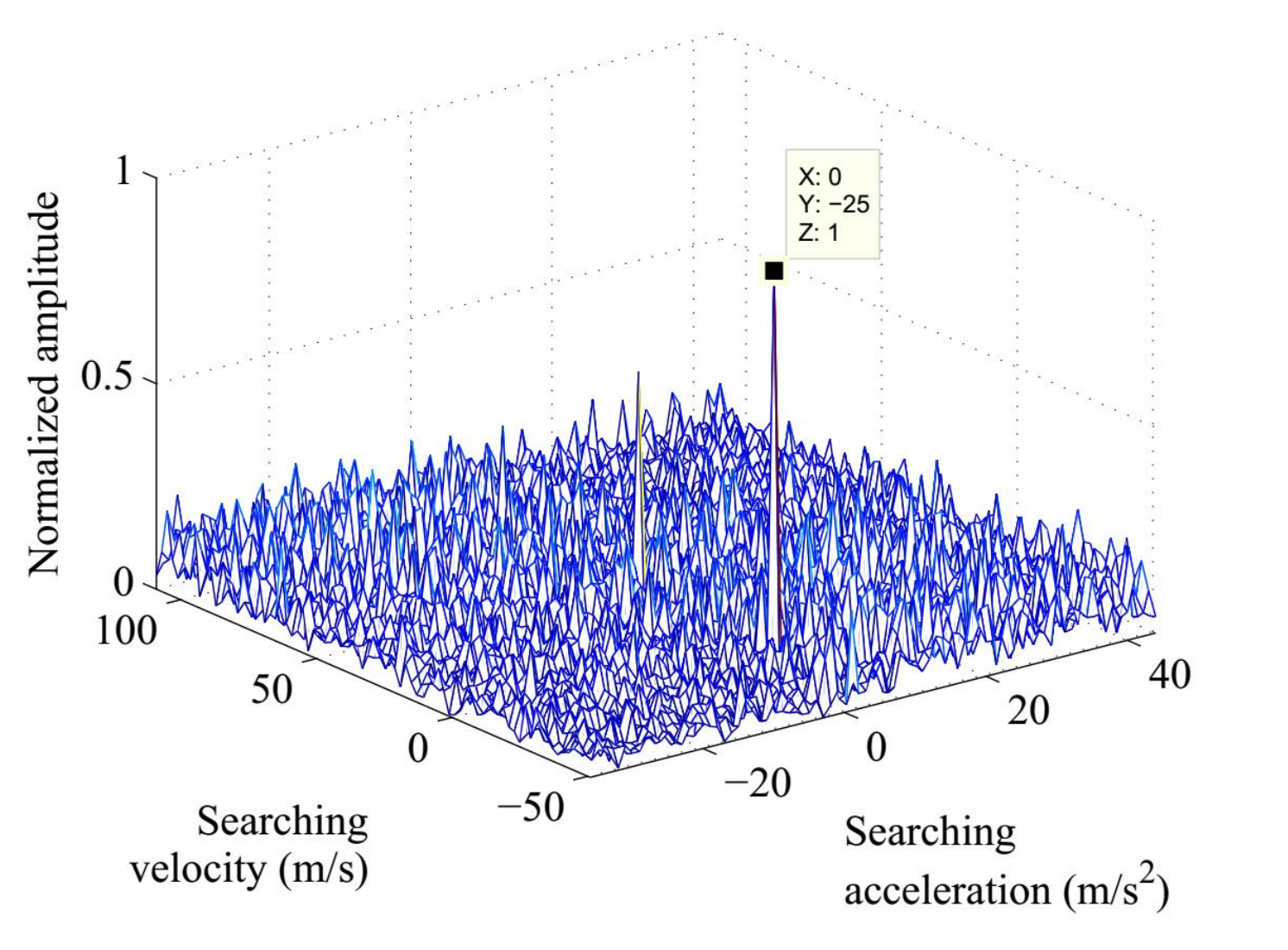}
\end{minipage}
}
\subfigure[]{
\begin{minipage}[b]{0.45\textwidth}\label{fig:real:e}
\includegraphics[width=1.\textwidth,draft=false]{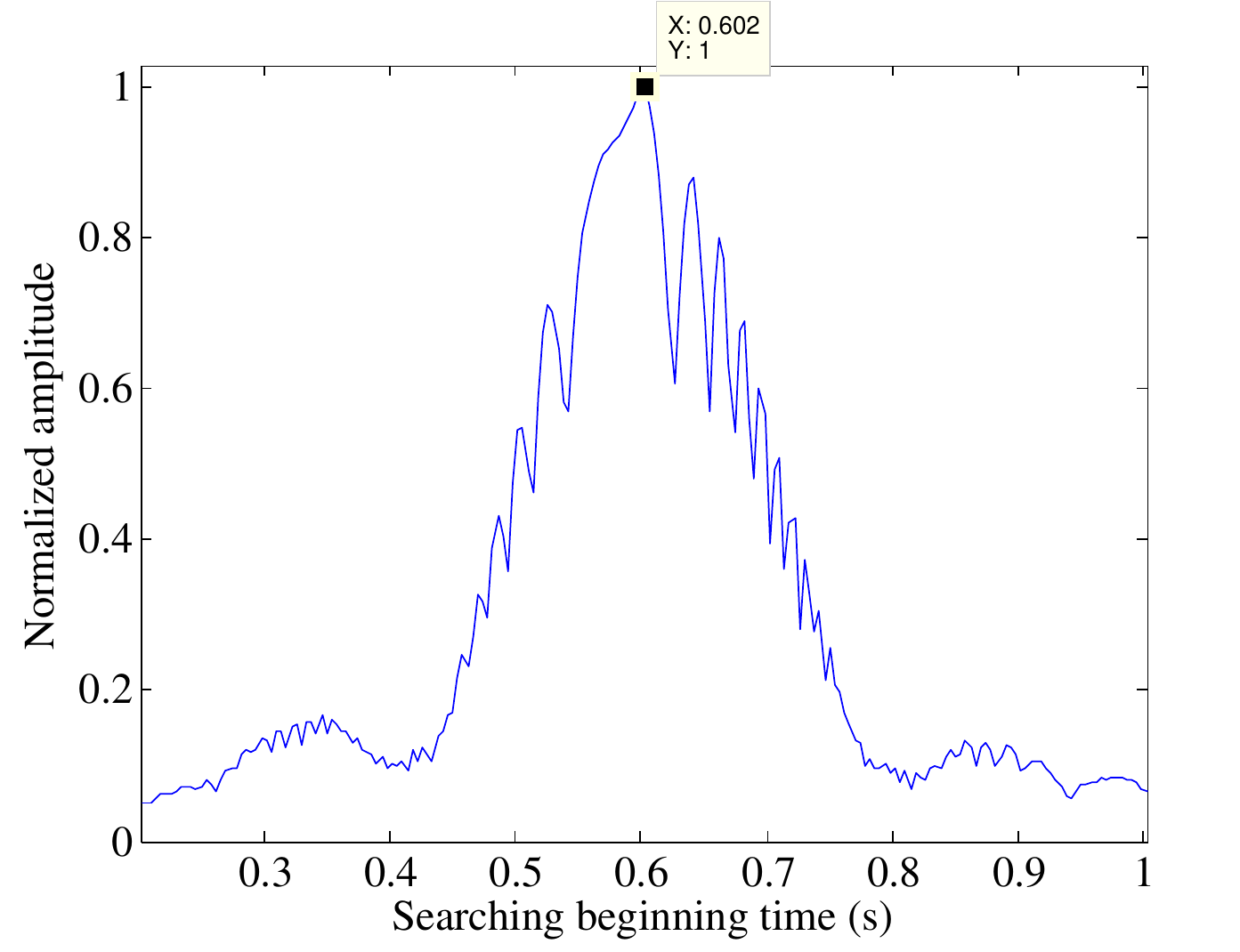}
\end{minipage}
}
\subfigure[]{
\begin{minipage}[b]{0.45\textwidth}\label{fig:real:f}
\includegraphics[width=1.\textwidth,draft=false]{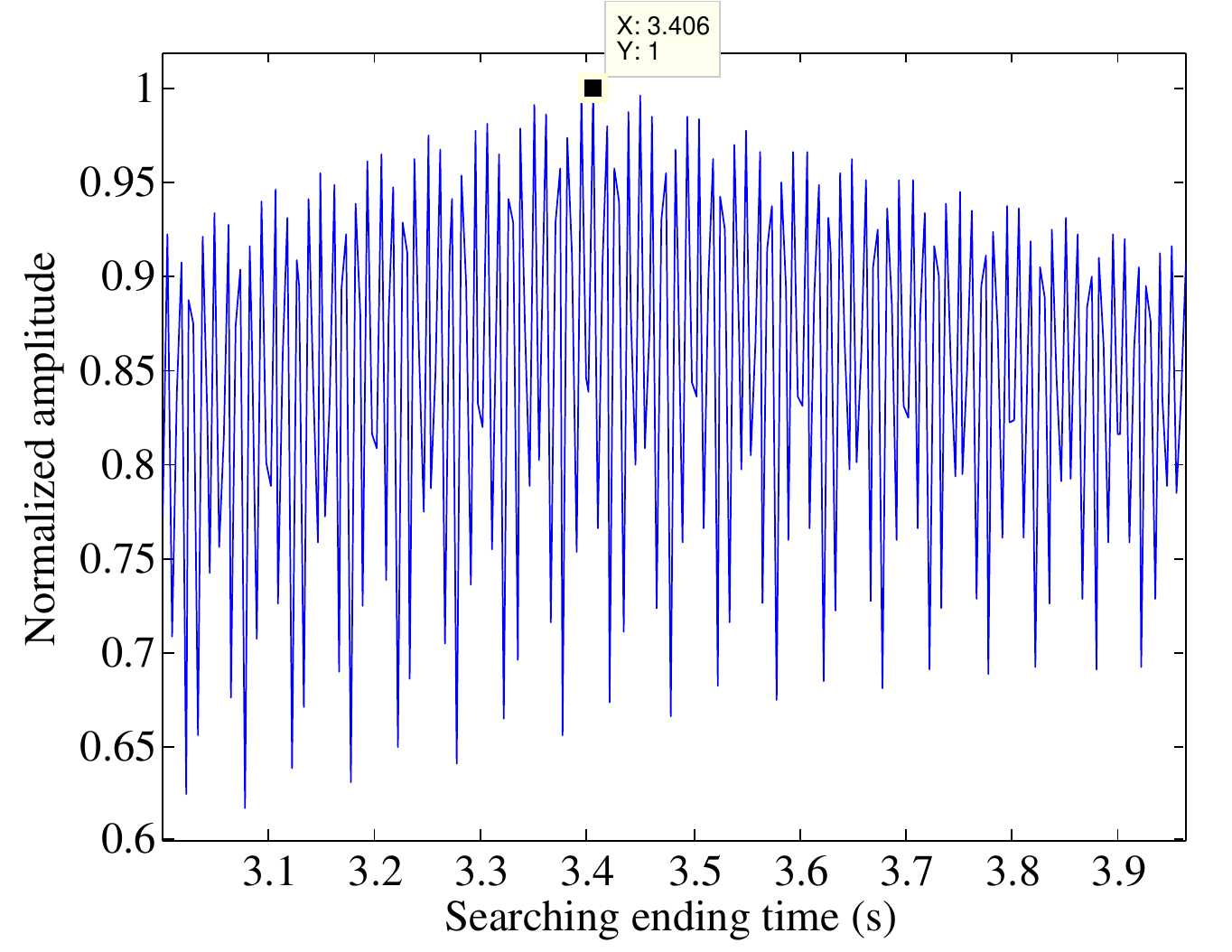}
\end{minipage}
}
\caption{Real data results. (a) Compressed echoes.
(b)  The focusing result of WRFRFT. (c) Integration result of
 RFRFT. (d) Integration result of GRFT. (e) Beginning time response slice.
 (f) Ending time response slice.} \label{fig:real}
\end{figure*}

The selected data to be processed consists of 2000 pulses
(i.e., coherent processing interval is 4 s), as show in Fig.
\ref{fig:real:a},
from which we could see that the target motion trajectory crosses
several range cells. In particular, we could choose "two points"
(as shown in Fig. \ref{fig:real:a}) from the target's
moving trajectory to obtain a rough estimation of target's velocity, i.e.,
$v=(79-72)\times2.5\text{m}/(1.428s-0.974s)=27.533 \text{m}/{s}$.
After WRFRFT, the focusing result of target is given in Fig. \ref{fig:real:b}. We could
notice that the target signal is focused as a peak
and its position represents the target's velocity and acceleration (i.e.,
$29\text{m}/{s}$ and $4\text{m}/{s^2}$).
For comparison, the processing results of RFRFT and GRFT are also given
in Fig. \ref{fig:real:c} and Fig. \ref{fig:real:d}, respectively.
For the focusing result of RFRFT (Fig. \ref{fig:real:c}), it could be noticed that
there is no significant peak in the RFRFT output and noted that the
"pseudo or false peak"  location is corresponding to $-9 \text{m}/{s}$ and $0 \text{m}/{s^2}$.
In addition, the "pseudo or false peak" location of GRFT (Fig. \ref{fig:real:d}) is corresponding to $-25 \text{m}/{s}$ and $0 \text{m}/{s^2}$. Therefore, both RFRFT and GRFT could not focus
the target signal correctly, resulting in estimation error
and even false alarm.
Furthermore,
the beginning-time response slice and ending-time response slice of WRFRFT are given
in Fig. \ref{fig:real:e} and Fig. \ref{fig:real:f}, respectively. From Fig. \ref{fig:real:e} and Fig. \ref{fig:real:f}, we could obtain the estimations
of target signal's  beginning/ending time, i.e.,
0.602 s and 3.406 s.

\begin{table}[ht]
\begin{center}
\begin{tabular}{|c|c|}
\multicolumn{2}{c}{\textbf{Table \textbf{4}}
}\\
\multicolumn{2}{c}{Radar Parameters for Real Data
}\\[5pt]
 \hline
Wave band  &  C
\\
 \hline
Bandwidth  & 20 MHz
\\
 \hline
Sample frequency  & 60 MHz
\\
 \hline
Pulse repetition frequency  & 500 Hz
\\
 \hline
Pulse duration  & 18 us
\\
 \hline
\end{tabular}
\end{center}
\end{table}

\section{Conclusion}\label{sec:conclusion}

In this paper, a new coherently focusing and parameters estimation method
(i.e., WRFRFT) has been proposed for a radar maneuvering target
with ARC and DS effects, where the times of the target appears in and leaves the radar detection area are unknown. By employing the window function and searching along the target moving trajectory, the echo signal of a maneuvering target could be well matched and focused as a peak in the WRFRFT domain and the target's time parameters (entry time and departure time) and motion parameters (range, velocity and acceleration) could be estimated via the peak location. The performances are validated by detailed experiments using simulated data and real data.

\end{document}